\documentclass[a4paper,superscriptaddress,11pt]{article}
\usepackage{authblk}
\usepackage{graphicx}
\usepackage{psfrag}

\title{The Diffeomorphism Constraint Operator in Loop Quantum Gravity}
\author[a,b]{Alok Laddha}
\author[b]{Madhavan Varadarajan}

\affil[a]{Institute for Gravitation and the Cosmos\\
 Pennsylvania State University,  University Park, PA 16802-6300, U.S.A} 
\affil[b]{Raman Research Institute\\ Bangalore-560 080, India}

\begin{document}

\maketitle

\thispagestyle{empty}

\let\oldthefootnote\thefootnote
\renewcommand{\thefootnote}{\fnsymbol{footnote}}
\footnotetext{Email:alok@gravity.psu.edu,madhavan@rri.res.in}
\let\thefootnote\oldthefootnote

\begin{abstract}
We construct the smeared diffeomorphism constraint operator at finite
triangulation from the basic holonomy- flux operators of Loop Quantum
Gravity, evaluate its continuum limit on the Lewandowski- Marolf
habitat and show that the action of the continuum operator provides 
an anomaly free representation of the Lie algebra of 
diffeomorphisms of the 3- manifold. Key features of our analysis include:
(i) finite triangulation approximants to the curvature, $F_{ab}^i$ of the
Ashtekar- Barbero connection which involve not only small loop
holonomies but also small surface fluxes as well as an explicit dependence
on the edge labels of the spin network being acted on
(ii) the dependence of the small loop underlying the holonomy on both the 
direction and magnitude of the shift vector field (iii) continuum constraint
operators which do {\em not} have finite action on the kinematic Hilbert space,
 thus implementing a key lesson from recent studies of 
parameterised field theory by the authors.

Features (i) and (ii) provide the first hints in LQG of a conceptual similarity
with the so called ``mu- bar'' scheme of Loop Quantum Cosmology.
We expect our work to be of use in the construction of an anomaly free 
quantum dynamics for LQG.

\end{abstract}

\section{Introduction}
The definition of the Hamiltonian constraint operator in Loop Quantum Gravity
(LQG) is far from  unique. The non-uniqueness stems from the tension between
the local nature of the Hamiltonian constraint and the non- local 
nature of some of the basic operators used in its construction. Specifically, 
the Hamiltonian constraint depends on the curvature  $F_{ab}^i$ of the Ashtekar- Barbero
connection whereas the basic connection dependent operators are holonomies of the 
connection around spatial loops. While, classically, the curvature can be obtained
through a limit of small loop holonomies wherein the small loop shrinks down to a point, 
quantum mechanically, the action of the corresponding holonomy operators in LQG
does not have the recquisite continuity for the limit to exist. Hence, one proceeds as follows \cite{qsd1}.
A triangulation of the 3 manifold is chosen and a finite triangulation approximant to the Hamiltonian
constraint is constructed from finite triangulation approximants to the local fields which 
comprise it. The exact Hamiltonian constraint is only obtained in the continuum limit of
infinitely fine triangulation. Due to the discontinuous action of the holonomy operators, it 
turns out that the continuum limit of the finite triangulation approximant to the constraint
(which we will refer to as the ``constraint at finite triangulation'') depends on the detailed choices
of holonomy approximants to $F_{ab}^i$ at finite triangulation. It follows that an issue of crucial 
importance is whether one can somehow restrict the choices of these holonomy approximants and
settle on a less ambiguous definition of the quantum dynamics of LQG.

There are broadly two aspects to the choice of holonomy approximants. The first is related to the choice of
representation used to evaluate the holonomy \cite{perez}, the second to the choice of small loop 
around which the holonomy is evaluated. The current state of art uses a fixed spin (usually spin half)
representation and a choice of little loop only restricted by the (slightly subjective) criterion of 
simplicity \cite{qsd1,lm,aajurekreview}. Unfortunately, calculations in  simpler contexts 
suggest that these choices may be physically inappropriate. On the one hand a detailed study of  
Parameterized Field 
Theory \cite{ppftham,ttham} suggests that the representation of the small loop holonomy be tailored to that 
of spin network edge it acts upon. On the other studies of isotropic
Loop Quantum Cosmology (LQC)\cite{improvedlqc} suggest that the specification of the size of the small loop
should involve the electric flux operator. Thus, not only is the definition of the Hamiltonian constraint operator
in LQG highly choice dependent, it may also be the case that the current set of choices are 
physically inappropriate.

In this work, we seek insight into the nature of the (possibly) correct set of choices through an analysis of the 
diffeomorphism constraint. The diffeomorphism constraint in LQG is handled very differently from the Hamiltonian 
constraint. LQG kinematics provides a unitary representation of finite spatial diffeomorphisms \cite{alm2t}. The 
diffeomorphism constraint is not imposed directly but, rather, by demanding that states be invariant under the action
of these unitary operators. Indeed, the (putative) generator of these unitary operators (which would correspond to the
diffeomorphism constraint) is not even defined on kinematic states because of the lack of weak continuity of 
the unitaries. 

Here, we treat the diffeomorphism constraint in a manner similar to the Hamiltonian constraint.
Accordingly, we fix a triangulation 
%(more precisely, a cubulation) 
of the 3- manifold and seek finite triangulation approximants to the 
various local fields which make up the diffeomorphism constraint, construct the diffeomorphism constraint at 
finite triangulation and then take its continuum limit. Since we know the action of {\em finite} diffeomorphisms,
our choice of finite triangulation approximants is guided by the requirement that the finite triangulation diffeomorphism
constraint have an action of the form 
\begin{equation}
-i\hbar\frac{{\hat U}_{\phi ({\vec N},\delta)}- {\bf 1}}{\delta}.
\label{final}
\end{equation}
Here $\delta$ parameterizes the fineness of the triangulation as well as the size of the diffeomorphism
$\phi ({\vec N},\delta )$ generated by the shift vector ${\vec N}$, ${\hat U}_{\phi}$ denotes the unitary 
operator corresponding to the diffeomorphism $\phi$ and the continuum limit is defined by $\delta\rightarrow 0$.
This requirement, while natural in itself, is also strongly motivated by our previous studies of PFT \cite{ppftham}.
Finally, we define the continuum limit of the finite triangulation diffeomorphism constraint operator on the 
Lewandowski- Marolf (LM) habitat \cite{lm} and check its physical appropriateness by showing that the algebra 
of diffeomorphism constraint operators on this habitat is anomaly free.

We shall see that the choice of approximant to the $N^aF_{ab}^i$ term in the diffeomorphism constraint reflects
similarities both with the choices made in PFT \cite{ppftham} as well as (at least at a conceptual level) LQC
\cite{improvedlqc}.

In the next section we outline the steps in our construction and detail the plan of the paper.

Before doing so, we would like to acknowledge the importance of early pioneering works on the diffeomorphism 
constraint in the context of the Loop Representation \cite{rs,tedlee} in providing inspiration and encouragement for the more
rigorous work done here, specifically the works of Bruegmann and Pullin \cite{berniejorge} and
Blencowe \cite{miles}.
%and Loll \cite{loll}. 
There, the diffeomorphism constraint
was shown to generate infinitesmal diffeomorphisms in quantum theory under a certain assumption of
regularity of the wave functions (which is violated in the current, rigorous formulation of the theory).
\footnote{For an analysis of the diffeomorphism constraint algebra in the 
context of calculations on a lattice, see \cite{loll}.}
It is also pertinent to mention Thiemann's
work \cite{qsd3} on the specific form of the diffeomorphism constraint which appears as the Poisson bracket 
between a pair of smeared density one Hamiltonian constraints. 
There too,  a factor which 
is the difference between a finite diffeomorphism unitary operator and the identity is obtained. 
\footnote{Due to the density one nature of the form of the constraint considered in \cite{qsd3}, there is no  
factor of $\delta^{-1}$. As seen in \cite{habitat2} and emphasized in our concluding section, the absence of
this factor yields a trivial action of the operator on the LM habitat.}
From the point of view
of our work here, the derivation \cite{qsd3} neglects certain contributions which are of ``order 1'' in the 
Hilbert space norm. 
Nevertheless, Thiemann's work is remarkable 
in that it represents the first attempt to tackle the diffeomorphism constraint in the modern formulation of LQG.

\section{Brief Sketch of Main Steps}
The purpose of   this section 
is to sketch the main steps in the construction so as to give the reader a  rough global view of the logic; 
the discussion will be schematic
and the reader should not be perturbed if it does not follow the sketch in detail. In this section we shall 
set $G=\hbar =c=1$. Since our discussion in this section is schematic, we shall further simplify our presentation
by choosing the Barbero- Immirizi
parameter $\gamma$ to be unity (only!) in this section. 
 
The diffeomorphism constraint $D({\vec N})$ is 
\begin{eqnarray}
D({\vec N}) &=& \int_{\Sigma} {\cal L}_{\vec N}A_a^i {\tilde E}^a_i \\
            &=& V ({\vec N}) - {\cal G}(N^cA_c^i) 
\end{eqnarray}
where
\begin{eqnarray}
 V ({\vec N}) &=& \int_{\Sigma}N^aF_{ab}^i {\tilde E}^b_i \label{defv}\\
{\cal G}(N^cA_c^i) &=& \int_{\Sigma}N^cA_c^i {\cal D}_a{\tilde E}^a_i .
\label{defg}
\end{eqnarray}
Here $\Sigma$ is the 3- manifold, $A_a^i$ is the Ashtekar Barbero connection, $F_{ab}^i$ is its curvature,
${\tilde E}^a_i$ is the densitized triad and ${\vec N}\equiv N^a$ is the shift vector field.

Let $T(\delta )$ be a 1 parameter family of triangulations
%\footnote{As mentioned earlier, and as we will see in detail in Section \ref{III}, $T(\delta)$ is in fact a 
%cubulation, however we losely refer to it as triangulation in this section.} 
of $\Sigma$ with the continuum limit 
being $\delta \rightarrow 0$ and let $D_{T(\delta )},V_{T(\delta )}, {\cal G}_{T(\delta )}$ be finite 
triangulation approximants to the quantities  $D, V, {\cal G} $ of the above equations. Thus 
$D_{T(\delta )},V_{T(\delta )}, {\cal G}_{T(\delta )}$ are expressions which yield $D, V, {\cal G} $
in the continuum limit.

For simplicity consider a (non- gauge invariant) spin network state consisting of a single edge $e$ with 
spin label $j$ so that the state is just the $m,n$  component of 
an edge holonomy $h_{e\; m}^{(j)\;\;n}$ of the (generalized) connection along the 
edge $e$ in the representation $j$, the indices $m, n$ taking values in the set $1,..,2j+1$.  
In what follows we shall supress some of these
labels and  denote the state simply by $h_e$.

From (\ref{final}) of section 1, our desired result is:
\begin{equation}
(1+ i\delta {\hat D}_{T(\delta )}) h_e = h_{\phi (\delta, {\vec N})\circ e}.
\label{desired}
\end{equation}
where, $\phi (\delta, {\vec N})\circ e$ is the image of $e$ by the diffeomorphism 
$\phi (\delta, {\vec N})$ which translates $e$ by an amount $\delta$ along the 
integral curves of the shift vector field $N^a$ (see Fig 1a).

\psfrag{e}{$e$}
\psfrag{f}{$\phi(\vec{N},\delta)$}
\psfrag{g}{$\phi(\vec{N},\delta)\circ e$}

\begin{center}
\includegraphics[scale=0.7]{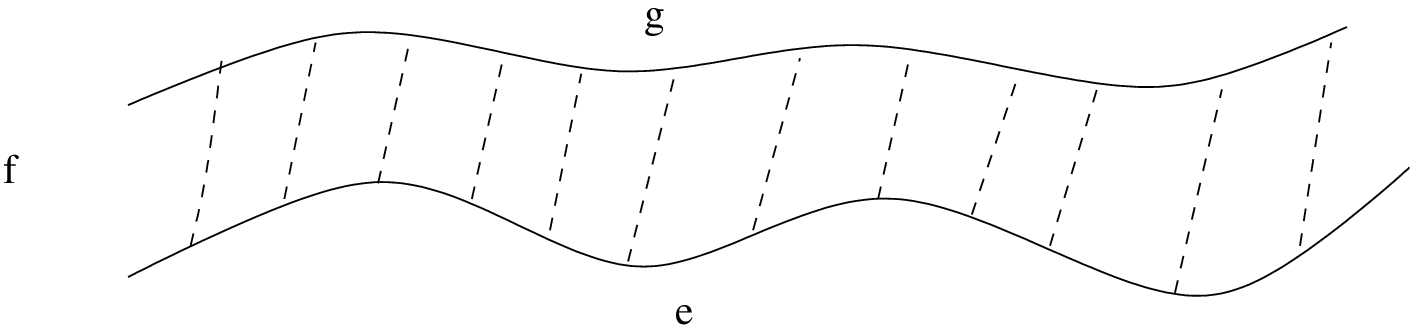}
Fig 1a
\end{center}

We obtain the desired result through the following steps:\\
\noindent (i) First we set 
\begin{equation}
(1+ i\delta \hat{ D}_{T(\delta )}):= (1+ i\delta \hat{{\cal G}}_{T(\delta )})(1+ i\delta {\hat V}_{T(\delta )}).
\end{equation}
\noindent (ii) Next, we show that
\begin{equation}
(1+ i\delta {\hat V}_{T(\delta )})h_e = h_{{\bar e}({\vec N}, \delta )}.
\label{bare}
\end{equation}
Here ${\bar e}({\vec N}, \delta )$ has the same end points as $e$ (as it must by virtue of the gauge 
invariance of $V$) and is obtained by joining the end points of 
$\phi (\delta, {\vec N})\circ e$ to those of $e$ by a pair of segments which are aligned with integral 
curves of $N^a$ as shown in Fig 1b.\\

\psfrag{m}{$e$}
\psfrag{n}{$\phi(\vec{N},\delta)$}
\psfrag{p}{$\overline{e}(\vec{N},\delta)$}

\begin{center}
\includegraphics[scale=0.7]{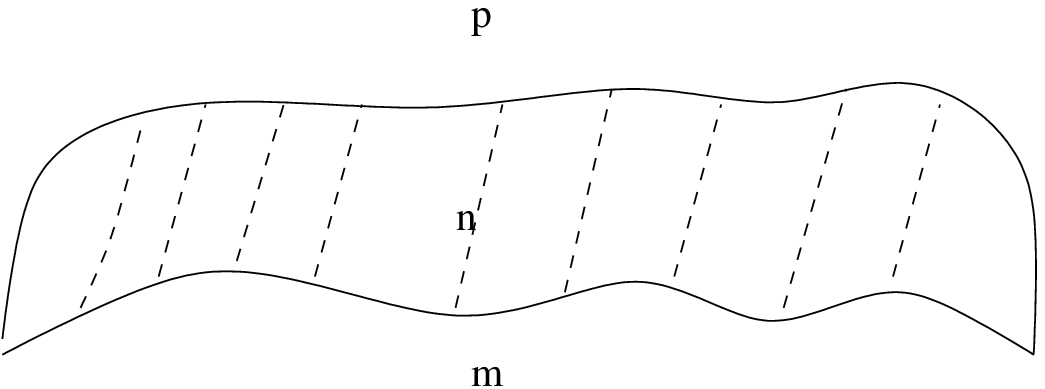}
Fig-1b
\end{center}

\noindent (iii) Finally,  we show that the 
Gauss Law term, $(1+ i\delta {\hat{\cal G}}_{T(\delta )})$ removes these two extra
segments (see Fig 1c).

\psfrag{r}{$\overline{e}(\vec{N},\delta)$}
\psfrag{s}{$\phi(\vec{N},\delta)\circ e$}

\begin{center}
\includegraphics[scale=0.7]{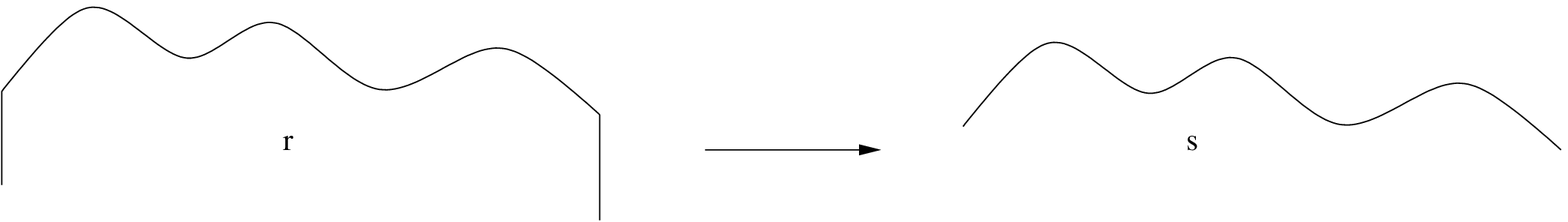}
Fig-1c
\end{center}

The major part of the analysis concerns the derivation of the identity (\ref{bare}) in step (ii) above.
We proceed as follows.

${ V}_{T(\delta )}$ is written as a sum over contributions $V_{\triangle}$ where $\triangle$ denotes a 
3- cell of the triangulation {\em dual} 
to $T (\delta )$, and $V_{\triangle}$ is a finite triangulation approximant to the integral
$\int_{\triangle}N^aF_{ab}^i {\tilde E}^b_i$. We order the triad operator to the right in ${\hat V}_{\triangle}$
so that only those 3- cells contribute which intersect $e$. 

The triangulation $T(\delta )$ is adapted to the edge $e$ so that its restriction to $e$ defines a triangulation of 
$e$. Thus, there is a triangulation of $e$ by 1- cells and vertices of $T(\delta )$ so that each of these vertices
$v_I, I=1,.., N$  is located at the centre of some 3- cell $\triangle = \triangle_I$. We define a finite triangulation
approximant to $F_{ab}^i$ in a such a way that the following identity holds:
\begin{equation}
(1+ i\delta {\hat V}_{\triangle_{I}})h_e = h_{{\bar e}(\triangle_I)}.
\end{equation}
$\forall\ I\ \in\ \{1,...,N-1\}$.\\
Here ${\bar e}(\triangle_I)$ is  obtained by moving the segment of $e$ between $v_I$ and $v_{I+1}$ along the integral 
curves of $N^a$ by an amount $\delta$ and joining this segment to the rest of $e$ at the points $v_I, v_{I+1}$
by a pair of segments which run along the integral curves of $N^a$  as shown in Fig 1d.

\psfrag{u}{$e$}
\psfrag{vI}{$v_{I}$}
\psfrag{vI+1}{$v_{I+1}$}
\psfrag{ebar}{$\overline{e}$}
\begin{center}
\includegraphics[scale=0.7]{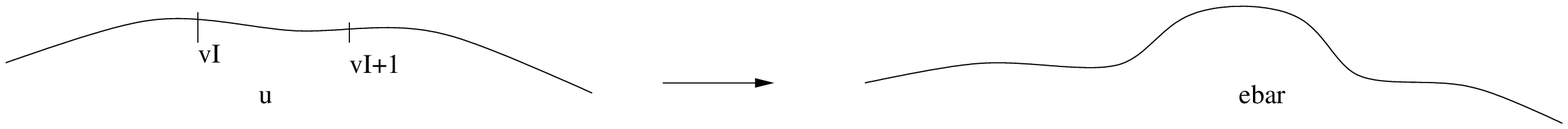}
Fig-1d
\end{center}

Next, we show that the contributions from all the ${\hat V}_{\triangle_I}$ yield the edge 
${\bar e}({\vec N}, \delta )$ of Fig 1b. Recall that ${ V}_{T(\delta )}$ is obtained by summing over all the cell
contributions $V_{\triangle}$. However,
summing over the action of all the ${\hat V}_{\triangle_I}$ on $h_e$ 
only yields a sum over states of the type $h_{{\bar e}(\triangle_I)}$.
In order to obtain the desired result, $h_{{\bar e}({\vec N}, \delta )}$, the sum over $\triangle$ is first converted
to a {\em product} over $\triangle$ i.e. to leading order in $\delta$, we show that 
\begin{equation}
1+ i\delta\sum_{\triangle} V_{\triangle} \sim \prod_{\triangle}(1+ i\delta V_{\triangle}).
\end{equation}
Hence, replacing  the sum over the corresponding operators by the product provides an equally legitimate 
definition of ${\hat V}_{T(\delta )}$. The replacement then leads, modulo some details, to the following identity
\begin{equation}
\prod_{\triangle}(1+ i\delta {\hat V}_{\triangle})h_e
= \prod_{I=1}^{N-1} (1+ i\delta V_{\triangle_I})h_e .
\end{equation}
Our definition of  ${\hat V}_{\triangle_I}$ is such that each factor
in the product acts independently, the $I$th factor acting only on the part of $e$ between $v_I$ and $v_{I+1}$.
We are then able to show that the result (\ref{bare}) follows essentially through the mechanism which is 
illlustrated schematically in Fig 1e.\footnote{The double lines in the figure indicate retraced paths.}

\psfrag{x}{$\overline{e}(\vec{N},\delta)$}

\begin{center}
\includegraphics[scale=0.7]{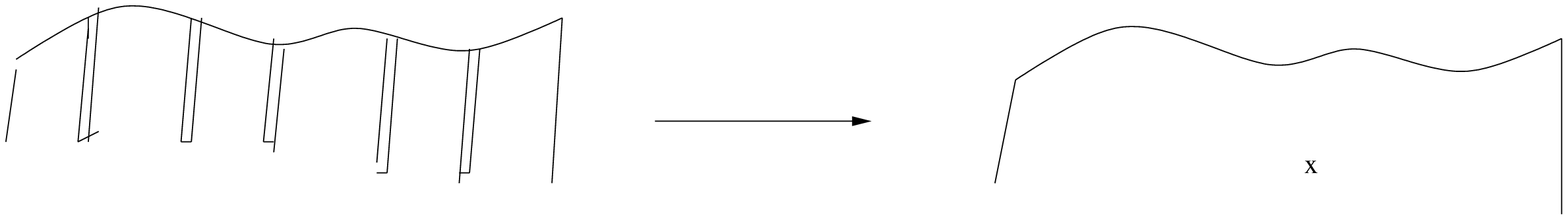}
Fig-1e
\end{center}

To summarise,  we obtain the desired result (\ref{desired}) through  the sequence: \\
Fig 1d$\rightarrow$Fig 1e$\rightarrow$Fig 1c.

We now turn to a description of the layout of the paper.
Sections \ref{III} to \ref{VI} are devoted to the case wherein the state is a (non gauge invariant) single edge spin network corresponding to an edge holonomy.
In section \ref{III} we describe our choice of triangulation. In section \ref{IV} we replace the finite triangulation in its natural form of a sum of over cells by a product over cells as advocated above and detail our choice of operator ordering of the 
product. Some associated technicalities are dealt with in appendix \ref{A1}. 
In section \ref{V} we construct finite triangulation approximants $V_{\triangle_I}$ from approximants to the 
curvature and triad. Section 5.1 is devoted to the spin half case and section 5.2 to the higher spin case.
Sections 5.1 and 5.2 furnish us with the key new results of this work: namely the approximants to the 
curvature terms. The treatement of the Gauss Law term ${\hat{\cal G}}_{T(\delta )}$ is relegated to the appendix
\ref{A2}. Since our main focus is on the curvature, our treatment of the Gauss Law terms is slightly heuristic.
In section \ref{VI} we put all the pieces together and obtain the desired expression (\ref{final}) 
for the diffeomorphism constraint operator.
Section \ref{VII} generalises the analysis of sections 3- 6 to the case of a general spin network.
Section 8 shows that the expression (\ref{final}) yields an anomaly free representation of the diffeomorphism
constraints on the Lewandowski- Marolf habitat.
Section 9 is devoted to a discussion of our results.

\section{Triangulation adapted to an edge $e$}\label{III}

Let $\Sigma$ be a real analytic, and compact (without boundary) oriented 3-manifold. Let $e$ be a 
closed, oriented, non-self intersecting, real analytic edge. Let $e$ be small enough that 
there exists an open neighbourhood ${\cal U}_{e}$ of $e$ with analytic chart $(x, y, z)$ such that the 
co-ordinate $x$ runs along $e$. 
\footnote{A general analytic edge can always be written as a composition of a finite number of such `small enough'
edges; the considerations of section 7 apply to such a composition.}

Let $\vec{N}$ be a real analytic vector field with affine parameter $\lambda$ 
so that $N^{a}\ =\ (\frac{\partial}{\partial\lambda})^{a}$. Let $\phi(\vec{N}, \delta)$ 
be the one parameter family of diffeomorphisms generated by $\vec{N}$ which translates any point 
in $\Sigma$ along an integral curve of $\vec{N}$ through affine length $\triangle\lambda\ =\ \delta$. 
Let $\delta$ be small enough  that $\phi(\vec{N},\delta)\circ e\ \subset\ {\cal U}_{e}$.\\
 Let $t_{e}$ be a triangulation of $e$ with 1-cells of co-ordinate length $\delta$. Let the vertices of $t_{e}$ be $v_{I}$ , $I=1,...,n$ with $v_{1}$ and $v_{N}$ being the beginning and end points of $e$ respectively.
Let $T$ be a triangulation
of $\Sigma$ such that $t_{e}\subset\ T$ so that $v_{I}$ are vertices of $T$. 
Let $T^{*}$ be a dual of $T$, such that each $v_{I}$ is in the interior of a 3-cell 
$\triangle_{I}$ of $T^{*}$.  Let the 3-cells $\triangle_{I}$ be cuboids of co-ordinate 
lengths ($\delta,\delta_{1},\delta_{2}$) along the ($x,y,z,$) directions, with $\delta_{1},\delta_{2}\ \ll\ \delta$.
We shall further require that $T^*$ be such that each vertex $v_I$ of $t_e$ be at the centroid (in the cordinates
$(x,y,z)$) of  the 3- cell $\triangle_{I}$.
Let  $T^*_e$ be the restriction of $T^*$ to $\cup_I \triangle_I$ i.e.
\begin{equation}\nonumber\\
T^{*}_{e}\ :=\ T^{*}\vert_{\cup_{I}\triangle_{I}}
\end{equation}

Next, let $\omega$ be a (smooth) three form on $\Sigma$. We restrict $T^*$ to be such that  every cell of $T^*$ 
which does not intersect $e$ 
has equal volume $v= \delta\delta_1\delta_2$ i.e. 
\begin{equation}
v\ =\ \int_{\triangle \notin T^*_e}\omega\ =\  \delta\delta_{1}\delta_{2}
\label{vol=v}
\end{equation}
Note that by
virtue of the smoothness of $\omega$, the analyticity of the chart $(x,y,z)$ and the compactness of
$T^*_e \subset \Sigma$, we have that 
\begin{equation}
\int_{\triangle_I}\omega  <  Dv,
\label{triangleibound}
\end{equation}
for some constant $D$ which is independent of $\delta, \delta_1,\delta_2, I$.

\section{Finite Triangulation Constraint:Sum to Product Reformulation}\label{IV}
In section 4.1 we recall the form of the continuum constraint and 
quantify the sense in which continuum quantities are approximated by finite triangulation ones.
The diffeomorphism constraint at finite triangulation 
appears naturally as a sum over cell contributions.
We recast this sum as a product over 3- cells of the triangulation in section 4.2
\footnote{\label{f41}The simple mechanism behind this reformulation is reflected in the identity 
$(\epsilon \sum_{i=1}^nx_{i} =\ \left(\prod_{i=1}^{n}(1 + \epsilon x_{i})\ -\ 1\right) + \textrm{O}(\epsilon^2)$
where $\epsilon$ is some small parameter and the $x_i$ are suitably bounded}
and specify the ordering of the corresponding  operator product in section 4.3. 

\subsection{Preliminaries}
The phase space of General relativity can be co-ordinatized with the (real) Ashtekar variables $(A_{a}^{i}, \tilde{E}^{b}_{j})$ with 
The Poisson bracket between the Ashtekar- Barbero connection $A_a^i$ and its conjugate densitized triad 
$\tilde{E}^{b}_{j})$ is 
\begin{equation}
\{A_{a}^{i}(x), \tilde{E}^{b}_{j}\}\ =\ (\gamma G)\delta^{3}(x,y)\delta_{a}^{b}\delta_{j}^{i}
\end{equation}
where $\gamma$ is the Barbero- Immirzi parameter.
The diffeomorphism constraint is:
\begin{equation}
\begin{array}{lll}
{ D}[\vec{N}]\ =\ \frac{1}{\gamma G}\int_{\Sigma} ({\cal L}_{\vec{N}}A_{a}^{i})\tilde{E}^{a}_{i} \\
\vspace*{0.1in}
\hspace*{0.5in}=\ \frac{1}{\gamma G}\int_{\Sigma} N^{a}F_{ab}^{i}\tilde{E}^{b}_{i}\ -\ \frac{1}{\gamma G}\int_{\Sigma} 
(N^aA_a^{i}){\cal D}_{b}\tilde{E}^{b}_{i}\\
\vspace*{0.1in}
\hspace*{0.5in}=\ \frac{1}{\gamma G}\{V[\vec{N}]\ -\ {\cal G}[ N^{i}]\}
\end{array}
\label{d=v+g}
\end{equation}
where $V[\vec{N}]$ is the vector constraint smeared with shift $N^a$ and ${\cal G}[N^i]$ is the  Gauss-constraint 
smeared with the connection dependent, $SU(2)$ Lie algebra valued Lagranage multiplier $N^i:=N^aA_a^i$.

In section 4.2 we shall approximate the continuum quantities defined above by 
quantities associated with the triangulation defined in section 3.
% $V_T({\vec N})$ by the replacement of certain cell contributions
%by their approximants. 
The error terms in the approximation vanish in the continuum limit of 
infinitely fine triangulation and hence are specified in terms of orders of the small parameters 
$\delta, \delta_1, \delta_2$ 
which measure the fineness of the triangulation. We shall use $X = X_T + O(\epsilon )$ to signify that
$\lim_{\epsilon\rightarrow 0} \frac{|X- X_T|}{\epsilon}$ exists.
If there is an additional parameter or index $p$ on which $X, X_T$ depend with  
$X(p)= X_T(p) + O(\epsilon)$, we shall say that 
$O(\epsilon )$ is independent of $p$ iff there exists a positive constant $C$ which is independent
of $p,\epsilon$  such that  for small enough $\epsilon$ we have that 
\begin{equation}
\frac{|X(p)- X_T(p)|}{\epsilon} < C
\label{orderindepp}
\end{equation}
We shall think of $\delta, \delta_1, \delta_2$ as independent parameters subject to $\delta_1,\delta_2 << \delta$
so that, for example, a quantity of order $O (\delta_1)$ is also of order $O(\delta )$. We shall often use
$v=\delta \delta_1\delta_2$ as a small parameter and its order is defined in terms of the primary parameters
$\delta, \delta_1,\delta_2$.

Note that if the length of the edge $e$ in the coordinate  $x$ is $L$,
we have that 
\begin{equation}
\sum_{I=1}^{N-1} \delta = L
\label{eq:0.1}
\end{equation}

\subsection{Product form of the classical constraint}
The vector constraint can be written as the following sum over cell contributions:
\begin{equation}\label{ap8-1}
V[\vec{N}]\ =\ \sum_{I=1}^{N}\int_{\triangle_{I}}N^{a}F_{ax}^{i}\tilde{E}^{x}_{i}\ +\ \sum_{I=1}^{N}\sum_{\hat{b}=y,z}\int_{\triangle_{I}}N^{a}F_{a\hat{b}}^{i}\tilde{E}^{\hat{b}}_{i} + \sum_{\triangle\notin T^{*}_{e}}\int_{\triangle}N^{a}F_{ab}^{i}\tilde{E}^{b}_{i}
\end{equation}

Next, let $V_{\triangle_{I}}^{(e)}$ be some function on the phase space such that
\begin{equation}\label{0}
|V_{\triangle_{I}}^{(e)} -\int_{\triangle_{I}}N^{a}F_{ax}^{i}\tilde{E}^{x}_{i}| < E\delta v
\label{vtrianglei}
\end{equation}
with $E$ being a postive number independent of $\delta,\delta_1,\delta_2 , I$.
Using equations (\ref{eq:0.1}),(\ref{vtrianglei}) in  
equation (\ref{ap8-1}), we have that 
\begin{equation}
V[\vec{N}]\ =\ \sum_{I}V_{\triangle_{I}}^{(e)} + \sum_{I}\sum_{\hat{b}=y,z}\int_{\triangle_{I}}N^{a}F_{a\hat{b}}^{i}\tilde{E}^{\hat{b}}_{i} + \sum_{\triangle\notin T^{*}_{e}}\int_{\triangle}N^{a}F_{ab}^{i}\tilde{E}^{b}_{i}\ +\ \textrm{O}(\delta)
\end{equation}
Using the mechanism outlined in Footnote \ref{f41},
and as shown in detail in
Appendix \ref{A1}, we recast  the `sum' form of $V[\vec{N}]$ into the `product' form:
\begin{equation}\label{eq:ap3-2}
\begin{array}{lll}
V[\vec{N}]\ =\\
\vspace*{0.1in}
\hspace*{0.3in}  \frac{-il_{p}^{2}\gamma}{\delta}\left\{\left[\prod_{I}\left(1 + \frac{\delta}{-il_{p}^{2}\gamma}V_{\triangle_{I}}^{(e)}\right)\right]\left[\prod_{I}\left(1 + \frac{\delta}{-il_{p}^{2}\gamma}\int_{\triangle_{I}}N^{a}F_{a\hat{b}}^{i}\tilde{E}^{\hat{b}}_{i}\right)\right]\right.\\
\vspace*{0.1in}
\hspace*{1.0in}\left.\left[\prod_{\triangle\notin T^{*}_{e}}\left(1 + \frac{\delta}{-il_{p}^{2}\gamma}\int_{\triangle}N^{a}F_{ab}^{i}\tilde{E}^{b}_{i}\right)\right]\ - 1\right\} + \textrm{O}(\delta)
\end{array}
\end{equation}
Equation (\ref{eq:ap3-2}) implies that the finite triangulation approximant $V_{T}[\vec{N}]$ to 
$V[\vec{N}]$ can be chosen as:
\begin{equation}\label{1}
\begin{array}{lll}
V_{T}[\vec{N}]\ =\\
\vspace*{0.1in}
\hspace*{0.3in}  \frac{-il_{p}^{2}\gamma}{\delta}\left\{\left[\prod_{I}\left(1 + \frac{\delta}{-il_{p}^{2}\gamma}V_{\triangle_{I}}^{(e)}\right)\right]\left[\prod_{I}\left(1 + \frac{\delta}{-il_{p}^{2}\gamma}\int_{\triangle_{I}}N^{a}F_{a\hat{b}}^{i}\tilde{E}^{\hat{b}}_{i}\right)\right]\right.\\
\vspace*{0.1in}
\hspace*{1.0in}\left.\left[\prod_{\triangle\notin T^{*}_{e}}\left(1 + \frac{\delta}{-il_{p}^{2}\gamma}\int_{\triangle}N^{a}F_{ab}^{i}\tilde{E}^{b}_{i}\right)\right]\ - 1\right\}
\end{array}
\label{defvtprod}
\end{equation}
The appearance of $l_{p}$ in a classical expression might seem unnatural. However,  
note that the continuum limit $\delta\rightarrow 0$ is distinct from the limit $\hbar\rightarrow 0$.
Equation (\ref{defvtprod}) reproduces the continuum classical function $V({\vec N})$ in the continuum limit
while keeping $\hbar$ (and, hence, $l_P$,) fixed.

Next, note that the diffeomorphism constraint of equation (\ref{d=v+g}) can be rewritten as the following product:
\begin{equation}
{D}[\vec{N}]\ =\ \frac{-i\hbar}{\delta}\left\{\left[1 - \frac{\delta{\cal G}[N^{i}]}{-il_{p}^{2}\gamma}\right]\left[1 + \frac{\delta V[\vec{N}]}{-il_{p}^{2}\gamma}\right]\ -\ 1\right\} + \textrm{O}(\delta)
\end{equation}
From equation (\ref{defvtprod}), it follows that the finite triangulation approximant to the diffeomorphism constraint,
$D_T({\vec N})$ can be chosen as:
\begin{equation}\label{eq:ap3-1}
{ D}_{T}[\vec{N}]\ =\ \frac{-i\hbar}{\delta}\left\{\left[1-\frac{\delta}{-il_{p}^{2}\gamma}{\cal G}_{T}[N^{i}]\right]\left[1 + \frac{\delta V_{T}[\vec{N}]}{-il_{p}^{2}\gamma}\right]-1\right\}
\end{equation}
where ${\cal G}_T$ is some finite triangulation approximant to ${\cal G}$.
%
%where the (Hamiltonian vector field associated with) ${\cal G}_{T}[N^{i}]$ is derived in Appendix \ref{A2}.\\
%We can now quantize ${ D}_{T}[\vec{N}]$ with the following operator ordering prescription.\\

\subsection{Operator ordering in quantum theory}
We define the operator, ${\hat D}_T({\vec N})$, corresponding to the approximant ${ D}_T({\vec N})$ of 
equation (\ref{eq:ap3-1})
through the following operator ordering prescription:\\
\noindent (i) Order the vector constraint piece to the right of the Gauss constraint piece as in (\ref{eq:ap3-1}).\\
\noindent (ii) Retain the order of the cell contributions in the product form of 
the vector constraint at finite triangulation as written in the classical expression (\ref{eq:ap3-2}).\\
\noindent  {(iii)} 
Order the densitized triad operators to the right in  each of the operators corresponding to the cell contributions 
$\int_{\triangle_{I}}{N^{a}F_{a\hat{b}}^{i}\tilde{E}^{\hat{b}}_{i}}$ , 
$\int_{\triangle\notin T^{*}_{e}}N^{a}{F_{ab}^{i}\tilde{E}^{b}_{i}}$.\\

The ordering (iii) ensures that both sets of operators discussed in (iii) annihilate the state $h_e$.
More in detail, since $e$ is in the x-direction,  terms containing $\hat{E}^{\hat{b}}_{i}$ do not contribute and
since $\triangle\cap e = \phi$ $\forall\ \triangle\ \notin T^{*}_{e}$, we have that
$\int_{\triangle} N^{a}\widehat{F_{ab}^{i}E^{b}_{i}}h_{e}\ =\ 0$. The ordering (i) and (ii) then yield the operator action
\begin{equation}\label{eq:ap3-3}
\begin{array}{lll}
\hat{{ D}}_{T}[\vec{N}]h_{e}\ =\\
\vspace*{0.1in}
 \hspace*{0.5in}\frac{-i\hbar}{\delta}\left\{\left[1-\frac{\delta}{-il_{p}^{2}\delta}\hat{{\cal G}}_{T}[N^{i}]\right]\left[\prod_{I}\left(1 + \frac{\delta}{-il_{p}^{2}\gamma}\hat{V}_{\triangle_{I}}^{(e)}\right)\right] - 1\right\}h_{e}
\end{array}
\end{equation}
Equation (\ref{eq:ap3-3}) can be re-written as
\begin{equation}\label{eq:ap3-4}
\begin{array}{lll}
1\ +\ \frac{\delta}{-i\hbar}\hat{{ D}}_{T}[\vec{N}]h_{e}=\\
\vspace*{0.1in}
\hspace*{0.6in} \left[1 - \frac{\delta}{-il_{p}^{2}\gamma}\hat{{\cal G}}_{T}[N^{i}]\right]\left[\prod_{I}\left(1 + \frac{\delta}{-il_{p}^{2}\gamma}\hat{V}_{\triangle_{I}}^{(e)}\right)\right]h_{e}
\end{array}
\end{equation}
As shown in appendix \ref{A2},
\begin{equation}
\left[1 - \frac{\delta}{-il_{p}^{2}}\hat{{\cal G}}_{T}[N^{i}]\right]\ =\ \hat{{\cal U}}^{SU(2)}_{(N^{i},\delta)}
\label{gtusu2}
\end{equation}
where $\hat{{\cal U}}^{SU(2)}_{(N^{i},\delta)}$ is the operator corresponding to the  finite $SU(2)$ 
(connection dependent) gauge transformation specified through the operator action
\begin{equation}
\hat{{\cal U}}^{SU(2)}_{(N^{i},\delta)} h_{e}\ =\ h_{s_{0}}^{-1}\cdot h_{e}\cdot h_{s_{1}}
\end{equation}
with $s_{0}$, $s_{1}$ being integral curves of $\vec{N}$ of parameter length $\delta$ 
at the beginning,final points of $e$  as detailed in appendix \ref{A2}.

Equations (\ref{eq:ap3-4}) and (\ref{gtusu2}) imply that the action of the 
finite triangulation diffeomorphism constraint operator  on a single edge holonomy is given by
\begin{equation}\label{eq:3}
\left(1 + \frac{\delta}{-i\hbar}\hat{ D}_{T}[\vec{N}]\right)h_{e}\ =\ \hat{{\cal U}}^{SU(2)}_{(N^{i},\delta)}\left[\prod_{I}\left(1 + \frac{\delta}{-il_{p}^{2}\gamma}\hat{V}_{\triangle_{I}}^{(e)}\right)\right]h_{e}
\end{equation}
In the next section we construct ${V}_{\triangle_{I}}^{(e)}$ and its operator correspondent, 
$\hat{V}_{\triangle_{I}}^{(e)}$.

\section{Curvature approximants}\label{V}

In this section we cosntruct cell- approximants to $N^{a}F_{ax}^{i}$
and $\tilde{E}^{x}_{i}$ in terms of the basic holonomy- flux variables. The approximant
${V}_{\triangle_{I}}^{(e)}$ is built out of these approximants and the operator
$\hat{V}_{\triangle_{I}}^{(e)}$ is defined by replacing the holonomy- flux variables in 
${V}_{\triangle_{I}}^{(e)}$ so constructed by their operator correspondents.

Recall that each vertex, $v_I$, of $t_e$ is located at the centre of the cell ${\triangle}_I$.
It follows that the rightmost 
face of $\triangle_{I}$ 
intersects $e$ at a point which is $\frac{\delta}{2}$ to the right of $v_I$.  
\footnote{ By rightwards, we mean in the direction of increasing `$x$'.} Let us call this face $S^{x}_{I}$ (this face is in the y-z plane by construction).
We approximate $\tilde{E}^{x}_{i}$ within $\triangle_{I}$ by, 
\begin{equation}\label{eq:1}
\begin{array}{lll}
\tilde{E}^{x}_{i}\ \approx\ \frac{\int_{S^{x}_{I}}\tilde{E}^{x}_{i}dy dz}{\int_{S^{x}_{I}}dy dz}\\
\vspace*{0.1in}
\hspace*{0.4in} = \frac{\int_{S^{x}_{I}}\tilde{E}^{x}_{i}dy dz}{\delta_{1}\delta_{2}}\\
\vspace*{0.1in}
\hspace*{0.1in} :=\ \frac{E_{i}(S^{x}_{I})}{\delta_{1}\delta_{2}}
\end{array}
\end{equation}
where $E_{i}(S^{x}_{I})$ is the electric flux through $S^{x}_{I}$.

By virtue of the compactness of $T^*_e\subset\Sigma$ and the smoothness of  
$\tilde{E}^{x}_{i}$
% and all it's derivatives are bounded within $\triangle_{I}$ it is easy to see that,
it follows that 
\begin{equation}\label{eq:ap4-1}
\tilde{E}^{x}_{i}({\vec x})\ =\ \frac{E_{i}(S^{x}_{I})}{\delta_{1}\delta_{2}}\ + {O}(\delta)
\end{equation}
$\forall\ {\vec x}\equiv (x,y,z)\ \in\ \triangle_{I}$ with the $O(\delta )$ term independent of $I,{\vec x}$ (see the 
discussion around equation (\ref{orderindepp})).

In sections (\ref{V.1}) and (\ref{V.2}) below we construct discrete approximants $(N^{a}F_{ax}^{i})_{I}$ to $N^{a}F_{ax}^{i}$. i.e. 
 \begin{equation}\label{eq:ap4-2}
 N^{a}F_{ax}^{i}({\vec x})\ =\ (N^{a}F_{ax}^{i})_{I}\ +\ {O}(\delta)
 \end{equation}
$\forall\ {\vec x}\equiv (x,y,z)\ \in\ \triangle_{I}$ with, as above, the  $O(\delta )$ term independent of $I,{\vec x}$. 
 Using (\ref{eq:ap4-1}), (\ref{eq:ap4-2}) we have,
 \begin{equation}\label{ap6-1}
 \int_{\triangle_{I}}N^{a}F_{ax}^{i}({\vec x})\tilde{E}^{x}_{i} d^{3}x\ =\ (N^{a}F_{ax}^{i})_{I}\ E_{i}(S^{x}_{I})\ \frac{v}{\delta_{1}\delta_{2}}\ +\ {O}(\delta v)
 \end{equation}
with  ${O}(\delta v)$ independent of $I$.
In conjuction with (\ref{0}) and the fact that $v = \delta\delta_{1}\delta_{2}$ the 
above equation implies that we may choose  $V_{\triangle_{I}}^{(e)}$ as
\begin{equation}\label{2}
 V_{\triangle_{I}}^{(e)}\ =\ (N^{a}F_{ax}^{i})_{I}E_{i}(S^{x}_{I})\delta
\end{equation}
We shall order the flux operator to the right (as in the classical expression above) in the quantum theory.
Since $S^{x}_{N}\cap e =\empty$, it follows that 
\begin{equation}
{\hat V}_{\triangle_{N}}^{(e)}h_e=0
\label{rightmostdualface}
\end{equation}
so that we may drop the 
rightmost $I=N$ factor in equation (\ref{eq:3}). 
  
Next, note that since $\vec{N}$ is real analytic and  $(x,y,z)$ is an analytic chart, it follows that either $N^{y},\ N^{z}$ both vanish at finitely many points on $e$ or $\vec{N}$ is along $e$. When $\vec{N}$ is along $e$, we have $(N^{a}F_{ax}^{i})_{I}\ :=\ (N^{x}F_{xx}^{i})_{I}\ =\ 0$ $\forall\ I$.\\
  It follows from (\ref{eq:3}) that in this case we only have the Gauss constraint term, which from the corollary in appendix \ref{A2}, yields the desired action (\ref{final}).
Hence it suffices to consider the case where $\vec{N}$ is transversal to $e$ except at finitely many points.  
Furthermore from the analysis of Section (\ref{VII}), it follows that, without loss of generality,
we can always choose the edge $e$ small enough so that $\vec{N}$ is transversal to $e$ everywhere 
except perhaps at it's end points.
 With the above restrictions on the edge $e$ and shift $\vec{N}$, 
%(which entails no loss of generality as we argued above) 
we construct $V_{\triangle_{I}}^{(e)}$ for a state which is a $j=\frac{1}{2}$ edge holonomy in section 5.1 and for 
a  $j >\frac{1}{2}$ edge holonomy in section 5.2. In both cases we show that 
$\hat{V}_{\triangle_{I}}^{(e)}$ acts in 
accordance with Figure 1d of section 2.

\subsection{The $j=\frac{1}{2}$ case}\label{V.1}

As is usually done, we shall approximate the curvature term $N^{a}F_{ax}^{i}$  to leading order in $\delta$ by 
a small loop holonomy. 
However, as we shall see, 
in order to obtain the desired action of $\hat{V}_{\triangle_{I}}^{e}$ , 
the small loop holonomy will be augmented by terms which are higher order in $\delta$.

Since we seek an approximant to the curvature in each cell $\triangle_I$, 
we shall associate a small loop $\gamma_{I}$
to each cell $\triangle_{I}\in T^*_{e}$ as follows. Let the part of the edge $e$ between (and including) the vertices
$v_I, v_{I+1}$ be $e_I$.
Let the image of $e_{I}$
under the small diffeomorphism $\phi_{\vec{N}, \delta}$ be $\phi_{\vec{N},\delta}\circ e_{I}$ with 
endpoints $\phi_{\vec{N},\delta}\circ v_{I}$ and $\phi_{\vec{N},\delta}\circ v_{I+1}$. 
Let the oriented segment between $v_{I}$ and $\phi_{\vec{N},\delta}\circ v_{I}$ be $s_{I}$, 
where the orientation is given by the direction of the vector field $\vec{N}$.   
Let the surface $S^x_I$ intersect $e_I$ at the point $v$. From our choice of triangulation as detailed in section 3,
it follows that 
the point $v$ is at the midpoint (as defined by the $x$ coordinate) of $e_I$.

Let $\overline{e}_{I}\ :=\  s_{I}\circ (\phi_{\vec{N},\delta}\circ e_{I})\circ s_{I+1}^{-1}$. Let the segment along $e$ connecting 
any two points $v^{'}$ with $v^{''}$ which lie in $e$ be $e_{v^{'},v^{''}}$.\footnote{This segment will be oriented along $e$ or opposite to $e$ depending on $v^{'}$ and $v^{''}$.}
The loop $\gamma_{I}$ is then defined as
\begin{equation}\label{3}
\gamma_{I}\ =\ e_{{v},v_{I}}\circ \overline{e}_{I}\circ e_{v_{I+1},{v}}
\end{equation}

We define the  curvature approximant $(N^{a}F_{ax}^{i})_{I}$ associated to the cube $\triangle_{I}$ as
\begin{equation}\label{eq:(4)}
(N^{a}F_{ax}^{i})_{I}\  :=\ \frac{1}{\delta^{2}}\{\ -Tr(h_{\gamma_{I}}\cdot \tau_{i})\ -\ \frac{2i}{3}Tr(h_{\gamma_{I}}-1)\frac{E_{i}(S_{I}^{x})}{\gamma l_{p}^{2}}\}
\end{equation}
Here Tr is trace in the fundamental representation. and $\tau_{i}\ =\ -i\sigma_{i}$, $\sigma_i, i=1,2,3$ 
being the Pauli matrices.

As shown in appendix \ref{A3}, the first term reproduces $N^{a}F_{ax}^{i}$ up to terms of $O(\delta)$. 
The second term is easily seen to be of higher order in $\delta$ because 
$Tr(h_{\gamma_{I}} - 1)\ \sim\ O(\delta^{3})$, and $E_{i}(S_{I}^{x})\ \sim\ O(\delta_{1}\delta_{2})$.
Thus equation (\ref{eq:(4)}) is a legitimate approximant to $N^aF_{ax}^i$. Nevertheless, its construction
has two striking features:\\
\noindent {(i)}
The role of 
the shift, which is a Lagrange multiplier ,in the construction of the small loop underlying the holonomy:
The traditional choice in LQG would be to construct small loops $\gamma_{ax}, a=y,z$ in the $a-x$ plane whose size
was independent of $N^a$ and have $N^a(v)$ appear as a multiplicative factor. Here, both the direction and {\em the 
magnitude} of $N^a$ are used in the construction of $\gamma_I$.\\
\noindent (ii) 
The unexpected electric flux dependence of an approximant to a purely connection dependent continuum quantity:
The traditional curvature approximant would only have the first term of equation (\ref{eq:(4)}) (modulo the 
comment (i) above). Moreover the second term blows up in the limit $\hbar\rightarrow 0$. Hence, as emphasized before,
it is imperative to take the continuum limit while keeping $\hbar$ fixed. While the $l_P^{-2}$ dependence disappears
in the continuum limit of the approximant to the classical curvature, it is conceivable that such a dependence
remains in the quantum theory (for a composite operator depending on the curvature) 
even after taking the continuum limit. If so, such factors, could concievably be the seeds
of non-perturbative quantum effects.\\

From (\ref{2}) and (\ref{eq:(4)}), we obtain
\begin{equation}
V_{\triangle_{I}}^{(e)}\ =\ \frac{-1}{\delta}\{ Tr(h_{\gamma_{I}}\tau^{i})\ +\ \frac{2i}{3}\frac{Tr(h_{\gamma_{I}}-1)E^{i}(S^{x}_{I})}{l_{p}^{2}\gamma} \}E^{i}(S^{x}_{I})
\end{equation}
Next, we evaluate the action of the corresponding quantum operator 
$\hat{V}_{\triangle_{I}}^{(e)}$  on a gauge-variant spin network defined by a single edge $e$ and $j=\frac{1}{2}$.
Note that
\begin{equation}
h_{eA}^{\;\;\;\;B}\ =\  (h_{e_{v_1,v_{I}}}\cdot h_{e_{I}}\cdot h_{e_{v_{I+1},v_N}})_{A}^{\;\;\;\;B}
\end{equation}
where $A,B$ are spinor indices ranging from 1 to 2.
Since ${v}\in e_{I}$, $\hat{E}^{i}(S_{I}^{x})$ acts only on the $h_{e_I}$ part of the state \cite{areaoprtr} and we get 
\begin{equation}
\hat{E}^{i}(S_{I}^{x})h_{eA}^{\;\;\;B}\ =\ (h_{e_{v_1,v_{I}}})_{A}^{\;\;\;\;C}\left(\frac{-i l_{p}^{2}\gamma}{2} (h_{e_{v_{I},v}})_{C}^{\;\;\;\;E}\tau_{i E}^{\;\;\;\;F}(h_{e_{v, v_{I+1}}})_{F}^{\;\;\;\;D}
(h_{e_{v_{I+1},v_N}})_{D}^{\;\;\;\;B} \right)
\end{equation}
Whence,
\begin{equation}\label{eq:mar19-1}
\begin{array}{lll}
[(\textrm{Tr}\ \hat{h}_{\gamma_{I}}\tau^{i})\hat{E}^{i}(S^{x}_{I})]h_{eA}^{\;\;\;B}\ =\\ 
\vspace*{0.1in}
\hspace*{0.3in} (h_{e_{v_1,v_{I}}})_{A}^{\;\;\;\;C}\left(\frac{-i l_{p}^{2}\gamma}{2}
(\ h_{\gamma_{I}M}^{\;\;\;\;N} (h_{e_{v_{I},v}})_{C}^{\;\;\;\;E}\tau_{i N}^{\;\;\;\;M}
\tau_{i E}^{\;\;\;\;F}(h_{e_{v, v_{I+1}}})_{F}^{\;\;\;\;D}\right)
(h_{e_{v_{I+1},v_N}})_{D}^{\;\;\;\;B} 
\end{array}
\end{equation}

Using the spinor identity:
\begin{equation}
\sum_{i}\ \tau_{i A}^{\;\;\;\;B}\ \tau_{i M}^{\;\;\;\;N}\ =\ [\delta_{A}^{\;\;\;\;B}\delta_{M}^{\;\;\;\;N}\ -\ 2\delta_{A}^{\;\;\;\;N}\delta_{B}^{\;\;\;\;M}],
\label{spinorid}
\end{equation}
it is easy to see that (\ref{eq:mar19-1}) can be simplified to:
\begin{equation}\label{eq:mar19-2}
\begin{array}{lll}
(\textrm{Tr}\ \hat{h}_{\gamma_{I}}\tau^{i})\hat{E}^{i}(S^{x}_{I})h_{eA}^{\;\;\;\;B}\ =\\ 
\vspace*{0.1in}
\hspace*{0.3in} \frac{-il_{p}^{2}\gamma}{2}\left(-2\ (h_{e_{v_1,v_{I}}}\circ h_{\overline{e}_{I}}\circ h_{e_{v_{I+1}},v_N})_{A}^{\;\;\;\;B}\ +\ 
(\textrm{Tr}\ h_{\gamma_{I}})h_{e A}^{\;\;\;\;B}\right)\\
\end{array}
\end{equation}
From (\ref{eq:mar19-1}), (\ref{spinorid}) we also have that 
\begin{equation}\label{eq:mar19-3}
\hat{E}^{i}(S_{x}^{I})\hat{E}^{i}(S_{x}^{I})h_{e A}^{\;\;\;\;B}\ =\ (-il_{p}^{2}\gamma)^{2}(\frac{-3}{4})h_{e A}^{\;\;\;\;B}
\end{equation}
Using (\ref{eq:mar19-2}) and (\ref{eq:mar19-3}), we obtain
\begin{equation}\label{eq:mar19-4}
\hat{V}_{\triangle_{I}}^{(e)}h_{e A}^{\;\;\;\;B}\ =\ \frac{(-il_{p}^{2}\gamma)}{\delta}\left((h_{e_{v_1,v_{I}}}\circ h_{\overline{e}_{I}}\circ h_{e_{v_{I+1}},v_N})_{A}^{\;\;\;\;B}\ -\ h_{e A}^{\;\;\;\;B}\right).
\end{equation}
Whence,
\begin{equation}
[\ 1+\frac{\delta}{(-il_{p}^{2}\gamma)}\hat{V}_{\triangle_{I}}^{(e)}\ ]h_{e A}^{\;\;\;\;B}\ =\ h_{\overline{e}_{\triangle_{I}} A}^{\;\;\;\;\:\:B}
\end{equation}
where $\overline{e}_{\triangle_{I}}$ is $e_{v_1,v_{I}}\circ \overline{e}_{I}\circ e_{v_{I+1},v_N}$.

This is the desired result (see equation (9) of section 2). \\

For future purposes, we note the following.
Since $\hat{E}^{i}(S^{x}_{I})$ acts non-trivially only on $e_{I}\subset\ e$ it follows that,
\begin{equation}\label{eq:mar19-5}
[\ 1\ +\ \frac{\delta}{(-il_{p}^{2}\gamma)}\hat{V}_{\triangle_{I}}^{(e)}\ ]h_{e_{I} A}^{\;\;\;\;\;B}\ =\ h_{\overline{e}_{I} A}^{\;\;\;\;\;B}
\end{equation}
Defining,
\begin{equation}\label{eq:mar25-4}
\begin{array}{lll}
\hat{f}_{1 I}^{i}\ =\ -(\textrm{Tr}\ h_{\gamma_{I}}\tau^{i})\\
\vspace*{0.1in}
\hat{f}_{2 I}\ =\ -\frac{1}{2i}(\textrm{Tr}\ h_{\gamma_{I}} -1),
\end{array}
\end{equation}
equation (\ref{eq:mar19-5})  can be rewritten as
\begin{equation}\label{eq:mar25-1}
\left[ 1 + \frac{\hat{f}_{1 I}\cdot\hat{E}(S^{x}_{I})}{(-i\gamma l_{p}^{2})}\ +\ \frac{\hat{f}_{2 I}\cdot\hat{E}(S^{x}_{I})\cdot\hat{E}(S_{I}^{x})}{\frac{3}{4}(-i\gamma l_{p}^{2})^{2}}\right]h_{e_{I} A}^{\;\;\;\;B}\ =\ h_{\overline{e}_{I} A}^{\;\;\;\;\;B}
\end{equation}
where we have used the notation $f\cdot g$ for $f_{i}g^{i}$.\footnote{Recall that the normalised Cartan killing metric on $su(2)$ is $\delta_{ij}$ which is used to define the scalar product on $su(2)$.}

\subsection{The $j>\frac{1}{2}$ case}\label{V.2}

Let $\pi_{j}[h_{e}]_{\alpha}^{\;\;\;\;\beta}$ be the holonomy along edge $e$ in spin-j representation 
so that $\alpha,\ \beta = \{1,...,2j+1 \}$.  Since the spin $j$ represenatation can be constructed as the symmetrized
product of  $n=2j$ copies of the fundamental representation, we have that 
\begin{equation}\label{mar25-11}
\pi_{j}[h_{e}]_{\alpha}^{\;\;\;\;\beta}\ =\ {\cal I}_{\alpha\ D_{1}...D_{N}}^{\beta C_{1}...C_{N}}h_{e C_{1}}^{\;\;\;\;D_{1}}...h_{e C_{N}}^{\;\;\;\;D_{N}}
\end{equation}
where the intertwining tensor ${\cal I}$ is symmetric under interchange of any of the $D$ indices 
(and hence also symmetric in the $C$ indices).
Our strategy for constructing the curvature approximant to $N^{a}F_{ax}^{i}$ is to first 
construct an operator $\hat{V}_{\triangle_{I}}^{(e)}$ such that
\begin{equation}\label{eq:11.6}
(1 + \frac{\delta}{-i\gamma l_{p}^{2}} \hat{V}_{\triangle_{I}}^{(e)})\pi_{j}[h_{e_{I}}]_{\alpha}^{\;\;\;\;\beta}\ =\ \pi_{j}[h_{\overline{e}_{I}}]_{\alpha}^{\;\;\;\;\beta}, 
\end{equation}
read off the expression for $\widehat{N^{a}F_{ax}^{i}}_I$ from that for  $\hat{V}_{\triangle_{I}}^{(e)}$ using
equation (\ref{2}) and show that the resulting expression is actually the operator correspondent of an approximant to
${N^{a}F_{ax}^{i}}_I$.
%\begin{equation}
%\hat{V}_{\triangle_{I}}^{(e)}\ =\ \widehat{(N^{a}F_{ax}^{i})}_{I}\hat{E}^{i}(S^{x}_{I})\delta
%\end{equation}

Two key observations which  help us construct $\hat{V}_{\triangle_{I}}^{(e)}$ are as follows.\\ 
\noindent {(1)} For sufficiently small $\delta_{1},\delta_{2}<<\delta$, it is easy to see that 
$S^{x}_{I}\cap {\bar e}_I= \empty$.
\begin{equation}\label{eq:key1}
\hat{E}^{i}(S^{x}_{I}) h_{\overline{e}_{I}}\ =\ 0.\\
\end{equation}
 \noindent {(2)} Since $\hat{E}(S^{x}_{I})\cdot\hat{E}(S^{x}_{I})$ is proportional to the Laplacian $\hat{J}^{2}$ on SU(2), we have
 \begin{equation}\label{eq:key2}
 \frac{\hat{E}(S^{x}_{I})\cdot\hat{E}(S^{x}_{I})}{(-i\gamma l_{p}^{2})^{2}}\pi_{k}[h_{e_{I}}]\ =\ k(k+1)\pi_{k}[h_{e_{I}}]
 \end{equation}
\\

Using the Leibnitz rule and the symmetry properties of ${\cal I}$ in (\ref{mar25-11})
we have that
\begin{equation}\label{eq:mar25-3}
 \hat{E}^{i}(S_{x}^{I})\pi_{j}[h_{e}]_{\alpha}^{\;\;\;\;\beta}\ =\ n{\cal I}_{\alpha\ D_{1}...D_{n}}^{\beta C_{1}...C_{n}}
 \left((\hat{E}^{i}(S^{x}_{I})h_{C_{1}}^{\;\;\;\;D_{1}})h_{C_{2}}^{\;\;\;\;D_{2}}...h_{C_{n}}^{\;\;\;\;D_{n}}\right)
\end{equation}
Next, using equations (\ref{eq:mar25-1}), 
(\ref{eq:mar25-4}), (\ref{eq:key2}) and (\ref{eq:mar25-3}) it is straightforward to show that
\begin{equation}
\left(1 +   \frac{\hat{f}_{1 I}\cdot\hat{E}(S^{x}_{I})}{n(-il_{p}^{2}\gamma)} + \frac{\hat{f}_{2 I}\hat{E}(S^{x}_{I})\cdot\hat{E}(S^{x}_{I})}{(-il_{p}^{2}\gamma)^{2}\frac{n}{2}(\frac{n}{2}+1)}\right)\pi_{j}[h_{e_{I}}]_{\alpha}^{\;\;\;\;\beta}\ =\ {\cal I}_{\alpha\ D_{1}...D_{n}}^{\beta C_{1}...C_{n}} \left( h_{\overline{e}_{I} C_{1}}^{\;\;\;\;D_{1}}h_{e_{I} C_{2}}^{\;\;\;\;D_{2}}...h_{e_{I} C_{n}}^{\;\;\;\;D_{n}} \right)
\end{equation}
Using (\ref{eq:key1}), the Leibnitz rule, the symmetry properties of ${\cal I}$, and equation (\ref{eq:key2})
for $\frac{1}{2}\leq k < j$ we iterate the above procedure to obtain,
\begin{equation}
\begin{array}{lll}
\prod_{m=1}^{n}\left(1 + \frac{\hat{f}_{1 I}\cdot \hat{E}(S^{x}_{I})}{m(-i\gamma l_{p}^{2})} + \frac{\hat{f}_{2 I}\hat{E}(S^{x}_{I})\hat{E}(S^{x}_{I})}{(-i\gamma l_{p}^{2})^{2}\frac{m}{2}(\frac{m}{2} + 1)}\right)\pi_{j}[h_{e_{I}}]_{\alpha}^{\;\;\;\;\beta}\ =\\
\vspace*{0.1in}
\hspace*{1.0in} {\cal I}_{\alpha\ D_{1}...D_{n}}^{\beta C_{1}...C_{n}}\left(h_{\overline{e}_{I} C_{1}}^{\;\;\;\;D_{1}}...h_{\overline{e}_{I} C_{n}}^{\;\;\;\;D_{n}}\right)\ =\ \pi_{j}[h_{\overline{e}_{I}}]_{\alpha}^{\;\;\;\;\beta}
\end{array}
\end{equation}
 Whence,  $\hat{V}_{\triangle_{I}}^{(e)}$ in (\ref{eq:11.6}) is given by
 \begin{equation}\label{eq:16}
 \delta\frac{\hat{V}_{\triangle_{I}}^{(e)}}{(-i\gamma l_{p}^{2})}\ =\ \prod_{m=1}^{n}\left(1 + \frac{\hat{f}_{1 I}\cdot\hat{E}(S^{x}_{I})}{m(-il_{p}^{2}\gamma)} + \frac{\hat{f}_{2 I}\hat{E}(S^{x}_{I}\cdot\hat{E}(S^{x}_{I})}{(-il_{p}^{2}\gamma)^{2}\frac{m}{2}(\frac{m}{2}+1)}\right)\ -\ {\bf 1}
\end{equation}

Expanding out the right hand side of (\ref{eq:16}), it is straightforward to see that the ``$-1$'' term is cancelled
by the ``$\prod_{m=1}^n1$'' term and that the remaining terms all have a factor of $\hat{E}(S^{x}_{I})$ to their right. 
Hence we may write equation (\ref{eq:16}) in the form:
\begin{equation}
\delta\frac{\hat{V}_{\triangle_{I}}^{(e)}}{(-i\gamma l_{p}^{2})} 
=: \hat{O}^{i}{\hat E}^{i}(S^{x}_{I})
\end{equation}

It remains to show that ${O}^i$ is a legitimate curvature approximant i.e. that  
\begin{equation}\label{eq:17}
O^{i}\ \approx\ \delta^{2}\frac{N^{a}F_{ax}^{i}}{(-i\gamma l_{p}^{2})} .
\end{equation}

A first analysis indicates that the (apparently) lowest order contribution (in $\delta$) to $O^i$ is the one which 
is independent of  $E^{i}(S^{x}_{I})$ i.e. the lowest order contribution to $O^i$ seems to be 
\begin{equation}
\sum_{m=1}^{n}\frac{f_{1 I}^{i}}{(-i\gamma l_{p}^{2})m}\ =\ \frac{f_{1 I}^{i}}{(-i\gamma l_{p}^{2})}\sum_{m=1}^{n}\frac{1}{m}.
\end{equation}
Since $\sum_{m=1}^{n}\frac{1}{m}\ >\ 1$ and since $\frac{f_{1 I}^{i}}{\delta^{2}}$ is a curvature approximant, it 
seems that $O^{i}$ is not a legitimate curvature approximant!

The solution to this apparent problem lies in the factors of $(-i\gamma l_{p}^{2})^{-1}$ in $\hat{O}^{i}$ which 
allow us to trade higher order holonomy-flux products by lower order ones by means of appropriate operator ordering. 
To see this, note that:
\begin{equation}\label{eq:18}
\frac{1}{(-il_{p}^{2}\gamma)}[\ \hat{E}(S^{x}_{I}), \hat{f}_{I 1}^{j}\ ]\ =\ \frac{1}{2}(\textrm{Tr}\ \hat{h}_{\gamma_{I}}\tau^{i}\tau^{j})
 \end{equation}
Viewed as a classical function, we have that 
\begin{equation}\label{eq:19}
 \frac{1}{2}(\textrm{Tr}\ h_{\gamma_{I}}\tau^{i}\tau^{j})\ =\ -\delta^{ij}\ +\ O(\delta^{2}).
\end{equation}
Thus, while an expression such as 
$\frac{\hat{f}_{I 1}\cdot \hat{E}(S^{x}_{I})\hat{f}_{I 1}^{j}}{(-il_{p}^{2}\gamma)}$ has the `naive' classical 
correspondent $\frac{f_{I 1}\cdot E(S^{x}_{I})f_{1 I}^{j}}{(-i l_{p}^{2}\gamma)}$ which is of 
$O(\delta_1\delta_2\delta^4)$, we may use 
 equation (\ref{eq:18}) to obtain: 
 \begin{equation}
 \frac{\hat{f}_{I 1}\cdot \hat{E}(S^{x}_{I})\hat{f}_{I 1}^{j}}{(-il_{p}^{2}\gamma)}\ =\ \frac{1}{(-il_{p}^{2}\gamma)}\hat{f}_{1 I}^{i}f_{1 I}^{j}E^{i}(S^{x}_{I})\ -\ \frac{1}{2}\hat{f}_{1 I}^{i}Tr(\hat{h}_{\gamma_{I}}\tau^{i}\tau^{j}),
\end{equation}
whose classical correspondent to leading order is seen to be  $f_{1 I}^{i}\ \sim\ O(\delta^{2})$ through equation 
(\ref{eq:19}).

Hence the question is whether theres exists some  operator ordering prescription with respect to which 
$\hat{O}^{i}$ may be identified with a finite triangulation approximant to the curvature term of interest.
We show below that answer to this question is in the afformative. Our prescription is as follows:
\noindent {(i)} Use the holonomy-flux commutation relation to move all the flux operators to the right.\\
\noindent {(ii)} Replace the holonomy-flux opeators in the resulting expression by their classical counter-parts.\\
\noindent {(iii)} Compute the leading order (in $\delta$) term in this expression. 
(From (\ref{eq:17}) we expect the resulting expression to be $O(\delta^{2})$.)

As stated above, $\hat{O}^{i}$ is defined through the equation
\begin{equation}\label{eq:20}
\hat{O}^{i}\hat{E}^{i}(S^{x}_{I})\ =\ \left[ \prod_{m=1}^{n}\left(1 + \frac{\hat{f}_{1 I}\cdot\hat{E}(S^{x}_{I})}{m(-i\gamma l_{p}^{2})} + \hat{f}_{2 I}\frac{\hat{E}(S^{x}_{I})\cdot\hat{E}(S^{x}_{I})}{(-i\gamma l_{p}^{2})\frac{m}{2}(\frac{m}{2}+1)}\right)\right] - 1
\end{equation}
Expanding the right hand side out yields an expression for ${\hat O}^i$ consisting of a sum of strings of holonomy and
flux operators. We apply step (i) to each of these strings. Note that the commutator between a flux operator and any
holonomy dependent term yields another holonomy dependent term. Hence the 
the classical correspondent of 
the commutator between a flux operator and any holonomy term is clearly at most of $O(1)$. This implies that any 
string with one or more flux operators located rightmost yields terms of $O(\delta_1\delta_2)$ ($\delta_1\delta_2$
being the area of the surface $S^{x}_{I}$). Choosing $\delta_1, \delta_2$ small enough, these terms are higher 
order than $\delta^2$ and hence may be ignored.

Hence, the only strings of interest in $\hat{O}^{i}$ are those which
%at the end of step (i),
%have no flux operators at all. Clearly such terms are obtained by the implementation of 
%are those 
%which (i) on strings 
which end in $\hat{f}_{1 I}^{i}$. We now show that even within this set of strings, 
strings which contain $\hat{f}_{2 I}\hat{E}(S^{x}_{I})\cdot\hat{E}(S^{x}_{I})$ yield only higher order
terms at the end of step (i).

\noindent
{\bf Lemma} : Any string in $\hat{O}^{i}$ (when it is expanded out as a 
sum of various composite operators) which ends in $\hat{f}_{1 I}$, 
but contains $\hat{f}_{2 I}\hat{E}(S^{x}_{I})\cdot\hat{E}(S^{x}_{I})$ is irrelevant.\\
\noindent
{\bf Proof} : \\ Note that $f_{2 I}$ is itself of O($\delta^{3}$). Hence $\hat{f}_{2 I}\hat{E}\cdot\hat{E}$ cannot be at the beginning of the string for an O($\delta^{2}$) contribution, as in that case no commutator can ``eat up" the $f_{2 I}$ term and the string will be at most of O($\delta^{3}$).  So consider the string of the type $[\hat{f}_{1 I}\cdot\hat{E}(S^{x}_{I})]^{m}\hat{f}_{2 I}\hat{E}(S^{x}_{I})\cdot\hat{E}(S^{x}_{I})[\hat{f}_{1 I}\cdot\hat{E}(S^{x}_{I})]^{n}\hat{f}_{1 I}^{i}$. As there is a $\hat{f}_{1 I}$ at the beginning point of the string, rest of the factors should (at the end of step (i)) conspire to yield an O(1) term .  However this is not possible as 
\begin{equation}
[\hat{f}_{2 I}, E^{j}(S^{x}_{I})]\ \alpha\ \textrm{Tr}(\hat{h}_{\gamma_{I}}\tau^{i})
\end{equation}
and $Tr(h_{\gamma_{I}}\tau^{i})$ is of O($\delta^{2}$).\\
Whence the operator string of the type $[\hat{f}_{1 I}\cdot\hat{E}(S^{x}_{I})]^{m}\hat{f}_{2 I}\hat{E}(S^{x}_{I})\cdot\hat{E}(S^{x}_{I})[\hat{f}_{1 I}\cdot\hat{E}(S^{x}_{I})]^{n}\hat{f}_{1 I}^{i}$ wil yield terms atmost of O($\delta^{4}$).\\
q.e.d.\\

We are now left only with strings which are of the form 
$(\hat{f}_{1 I}\cdot\hat{E}(S^{x}_{I}))^{n}\hat{f}_{1 I}^{i}$.
It is important to note that due to the leftmost occurance of $\hat{f}_{1 I}^{i}$, 
such strings can atmost be of O($\delta^{2}$). Hence we are only interested in contributions from 
the remaining part of the string (except the left most $\hat{f}_{1 I}$) which
give rise to terms of O(1) at the end of steps (i) and (ii). In particular if any terms at the end of step (i)
have any flux operators at all, they will be of higher order. So we only seek contributions
which, at the end of the application of step (i) to the strings of interest, are {\em independent} of the flux 
operators. We show below such contributions combine to yield 
a legitimate curvature approximant.

The relevant strings are of the form $(\hat{f}_{1 I}\cdot\hat{E}(S^{x}_{I}))^{n}\hat{f}_{1 I}^{i}$ and can be read off by expanding (\ref{eq:20}).\footnote{Equation (\ref{eq:20}) 
defines $\hat{O}^{i}\hat{E}^{i}(S^{x}_{I})$. The relevant terms we are analysing above are those terms which have $\hat{f}_{1 I}^{i}E^{i}(S^{x}_{I})$ sitting at the rightmost position in the string.} It is straightforward to read off the co-efficients of each relevant string from (\ref{eq:20}).
\begin{equation}\label{eq:mar28-1}
\begin{array}{lll}
c_{p}\ :=\ \textrm{Co-efficient of} \frac{(\hat{f}_{1 I}\cdot\hat{E}(S^{x}_{I}))^{p-1}\hat{f}_{1 I}^{i}}{(-i\gamma l_{p}^{2})}\\
\vspace*{0.1in}
\hspace*{0.3in} = \sum_{m_{p}=p}^{n}\sum_{m_{p-1}=p-1}^{m_{p}-1}...\sum_{m_{2}}^{m_{3}-1}\sum_{m_{1}=1}^{m_{2}-1}\frac{1}{m_{p}...m_{1}}\\
\vspace*{0.1in}
\hspace*{0.3in} =:\ \sum_{m_{p}>m_{p-1}>...>m_{1}}S_{m_{p}...m_{1}}
\end{array}
\end{equation}
Whence the sum of all the relevant strings (henceforth denoted as $\hat{T}$) is given by,
\begin{equation}
\begin{array}{lll}
\hat{T}\ =\ \sum_{p=1}^{n}c_{p}\left(\hat{f}_{1 I}\cdot\hat{E}(S^{x}_{I})\right)^{p-1}\hat{f}_{1 I}^{i}\\
\vspace*{0.1in}
\hspace*{0.3in}\ =\ \sum_{p=1}^{n}\sum_{m_{p}>...>m_{1}}S_{m_{p}...m_{1}}(\hat{f}_{1 I}\cdot\hat{E}(S^{x}_{I}))^{p-1}\hat{f}_{1 I}^{i}
 \end{array}
 \end{equation}
 Now we are in a position to apply steps (i), (ii) and (iii) to $\hat{T}$.\\
 Each string of the type $(\hat{f}_{1 I}\cdot\hat{E}(S^{x}_{I}))^{p}\hat{f}_{1 I}^{i}$ has $p$ intermediate factors of the type $\hat{E}^{j}(S^{x}_{I})\hat{f}_{1 I}^{k}$. As mentioned above, 
we are interested only in those terms which after step (i) have no
factors of the flux operator. Clearly for each such string there is precisely 1 such term which is obtained by
using the 
%Applying step (i) to such a string amounts to applying 
the commutator (\ref{eq:18}) p times to remove all the flux operators. \\
 %That is, as before we replace 
% $\hat{E}(S^{x}_{I})^{j}\hat{f}_{1 I}^{k}$ with $\left([\hat{E}(S^{x}_{I})^{j}, \hat{f}_{1 I}^{k}]\ =\ 2(-il_{p}^{2}\gamma)\textrm{Tr}(\hat{h}_{\gamma_{I}}\tau^{i}\tau^{j})\right)$ + $\hat{f}_{1 I}^{k}E^{j}(S^{x}_{I})$.%
Now applying step (ii) amounts to using (\ref{eq:19}) and ignoring the O($\delta^{2}$) term.  
This would mean that $\hat{T}$ transits to a classical quantity whose leading order part, $T$, is  given by:
\begin{equation}\label{eq:21}
\hat{T}\ =\ f_{1 I}^{i}\Big(\sum_{m=1}^{n}S_{m}\ -\ \sum_{m_{2}=1}^{n}\sum_{m_{1}=1}^{m_{2}}S_{m_{2},m_{1}}+...+\sum_{m_{n}>...>m_{1}}(-1)^{n-1}S_{m_{n}...m_{1}}\Big)
\end{equation}
Now note that, $\sum_{m_{2},m_{1}=1}$ cancels all the terms in $\sum_{m=1}^{n}S_{m}$ except $S_{1}$. Similarly, $\sum_{m_{2}<m_{3}}S_{m_{3},m_{2},m_{1}=1}$ cancels all the terms in $\sum_{m_{2}}S_{m_{2},m_{1}\neq 1}$ and so forth. \emph{Thus the term inside the bracket in (\ref{eq:21}) adds up to 1}!
This shows that to leading order in $\delta$ the classical quantity $O^{i}$ obtained through steps (i)- (iii)
%by naively replacing holonomy and flux operators by their classical counterparts 
equals $f_{1 I}^{i}$ and is, therefore, a legitimate approximant to curvature.

\section{Final operator expression for the Diffeomorphism constraint}\label{VI}

 As mentioned in the beginning of (\ref{V}), if $\vec{N}$ is along $e$, (\ref{eq:3}) in conjuction with the corollary in appendix \ref{A2} yields the desired action.
 
 \begin{equation}\label{22}
 \left(1 + \frac{\delta\hat{D}_{T}[\vec{N}]}{-i\hbar}\right)h_{e A}^{\;\;\;\;B}\ =\ 
h_{\Phi(\vec{N},\delta)\circ e A}^{\;\;\;\;\;\;\;\;\;\;\;\;\;\;\;\;\;\;B}
 \end{equation}

We will now argue that (\ref{22}) also holds when $\vec{N}$ is transverse to $e$ everywhere except perhaps at it's end points. (This is the same scenario, we worked with in section (\ref{V.1}), and section (\ref{V.2}).)\\
Notice that since,\\
\noindent{\bf (i)}
\begin{equation}\label{23}
\begin{array}{lll}
\hat{E}_{i}(S^{x}_{I})h_{e_{J}}\ =\ 0\ \forall I\neq J,\ I,\ J\ \in\ \{1,...,N-1\}\\
\vspace*{0.1in}
\hat{E}_{i}(S^{x}_{I})h_{\overline{e}_{J}}\ =\ 0, \forall\ I,\ J\ \in\ \{1,...,N-1\}
\end{array}
\end{equation}
and,\\
\noindent{\bf (ii)}
\begin{equation}\label{24}
h_{e A}^{\;\;\;\;B}\ =\ h_{e_{1} A}^{\;\;\;\;\;\;\;\;A_{1}}...h_{e_{N-1} A_{N-2}}^{\;\;\;\;\;\;\;\;\;\;\;\;\;\;\;\;\;B}
\end{equation}
where in the second line in (\ref{23}) we have used $\delta_{1},\delta_{2}\ \ll\ \delta$.
Equations (\ref{23}), (\ref{24}) together with (\ref{eq:mar19-5}), (\ref{eq:11.6})  and the fact that $\hat{E}_{i}(S^{x}_{I})$ is ordered to the right in $\hat{V}_{\triangle_{I}}^{(e)}$ imply that,
\begin{equation}\label{25}
\begin{array}{lll}
\prod_{I=1}^{N}\left[1 + \frac{\delta}{-il_{p}^{2}\gamma}\hat{V}_{\triangle_{I}}^{(e)}\right] h_{e A}^{\;\;\;\;B}\ &=&\ 
\prod_{I=1}^{N-1}\left[\left(1 + \frac{\delta}{-il_{p}^{2}\gamma}\hat{V}_{\triangle_{I}}^{(e)}\right)h_{e_{I} A_{I-1}}^{\;\;\;\;\;\;A_{I}}\right]\\
\vspace*{0.1in}
& & = \prod_{I=1}^{N-1} h_{\overline{e}_{I} A_{I-1}}^{\;\;\;\;\;A_{I}}\\
\vspace*{0.1in}
& & = h_{\overline{e}(\vec{N},\delta)}
\end{array}
\end{equation}
where we have set $A_{0} = A, A_{N-1}=B$ and have used the definition of $\overline{e}_{I}$ as given in  section (\ref{V.1}) (above eq. (\ref{24})) to define $\overline{e}(\vec{N},\delta)$.
\begin{equation}
\overline{e}(\vec{N},\delta)\ =\ s_{0}\circ\left(\phi(\vec{N},\delta)e\right)\circ s_{1}^{-1}
\end{equation}
Finally equations (\ref{3}), (\ref{25}) and (\ref{ap9-1}) in appendix \ref{A2} implies the desired result given in eq. (\ref{22}).

\section{The case of a general spin-network state}\label{VII}

In this section we generalize the considerations of sections 3- 6 to the case of a spin network based on a 
graph $\gamma$ with $M$ edges, $e_p, p\ =\ 1,..,M$. Our arguments closely parallel those in sections 3-6 and, as 
a result, our presentation will not be as detailed as in those sections.
Without loss of generality, we assume that each edge $e_p$ is of the type described in section \ref{III}. 
We shall also assume, (once again without loss of generality) that the shift vector field is 
along the edge $e_p$ or is transverse to it, except perhaps at its endpoints.

We denote the spin-network state by $\vert {\bf s}\rangle := \vert \gamma, \{ \vec{c}, \vec{j} \}\rangle $, where
${\vec j}, {\vec c}$ refer to the set of edge labels and intertwiners associated with the spin network.

\subsection{Triangulation}\label{VII.1}

We use an obvious generalization of the notation used in section 3. Let $e_p\subset {\cal U}_{p}$ where ${\cal U}_{p}$
is an open set equipped with analytic coordinates $\left(x_{p}, y_{p}, z_{p}\right)$ with $x_p$ running along $e_p$.
Let $\delta$ be small enough that $\phi({\vec N},\delta )\circ e_p \subset {\cal U}_{e_p}$.

Let $t_{e_p}$ be a triangulation of $e_p$ with vertices $v^{(p)}_{I_{p}}$, $I_{p}\ =\{1,...,N_{p}\}$ and 1- cells
of coordinate length $\delta$. Let $T^p$ be a triangulation of $\Sigma$ such that $t_{e_p}\subset T^p$.
Let $T^{* p}$ be dual to $T^{p}$ such that every vertex $v^{(p)}_{I_{p}}$ lies in the interior of 
some 3-cell $\triangle_{I_{p}}^{(p)}$ of $T^{* p}$.\\
Let the 3-cells $\triangle^{(p)}_{I_{p}}$ be cuboids with co-ordinate lengths $(\delta, \delta_{1},\delta_{2})$ 
along the $(x_{p}, y_{p}, z_{p})$ directions with $\delta_{1},\delta_{2}\ <<\ \delta$. 
We shall further require that the each vertex $v^{(p)}_{I_{p}}$ be located at the (coordinate) centroid of the 
3- cell $\triangle^{(p)}_{I_{p}}$.
Let $T^{*}_{e_{p}}$ be the restriction of $T^{* p}$ to $\cup_{I_{p}}\triangle_{I_{p}}^{(p)}$. 

Define  the set $T^{*}_{\gamma}$ by   
\begin{equation}
T^{*}_{\gamma}:=\cup_{p=1}^{M}T^{*}_{e_{p}}.
\label{t*gamma}
\end{equation}
$T^{*}_{\gamma}$ defines a subset of  $\Sigma$ (namely the union of all the 3- cells contained in
$T^{*}_{e_{p}}, p=1,..,M$))
which we call ${\cal U}_{\gamma}$.\\
Let $T^{* '}_{\gamma}$ be a triangulation of $\overline{\Sigma - {\cal U}_{\gamma}}$ 
with 3-cells $\triangle$ of volume $v=\delta\delta_{1}\delta_{2}$ i.e.
\begin{equation}
\int_{\triangle \in T^{* '}_{\gamma}} \omega =v
\label{volt*'gamma}
\end{equation}
where $\omega$ is the 3- form of section 3.
Note also that by virtue of the smoothness of $\omega$, the compactness of $T^*_{e_{p}}\subset \Sigma$
and the fact that the graph $\gamma$ has a finite number of edges, it follows that 
\begin{equation}
\int_{\triangle_{I_{p}}^{(p)}}\omega < Dv
\end{equation}
for some constant $D$ which is independent of $\delta, \delta_1,\delta_2, {I_{p}},p$.

Finally, note that $T^{*}_{\gamma}$ is not, strictly speaking, a triangulation of ${\cal U}_{\gamma}$ because some of its
3- cells overlap, namely the ones in the vicinity of the endpoints of the edges $e_p$ (recall that these 
endpoints are the vertices of the graph $\gamma = \cup_p e_p$).
However, it is easy to see that 
for small enough $\delta_1,\delta_2$ this overlap at any such graph vertex involves at the most
one 3- cell from each $T^{*}_{e_p}$ and yields a negligible ``overcounting'' error in terms of the approximation of
an integral of fields over ${\cal U}_{\gamma}$ by sums over 3- cells of $T^{*}_{\gamma}$.

We use this fact (that the contributions of such cells to the evaluation of such integrals are negligible) 
to remove, by hand, the contributions from the right most cells of each $T^{*}_{e_p}$ i.e. from the cells
$\triangle_{{N_p}}^{(p)}, p=1,..,M$ so that, in what follows, we shall allow $I_p$ to range from $1$ to $N_p-1$.
\footnote{
This removal can be further justified in the case when two edges which meet
at a vertex of $\gamma$ are analytic continuations of each other and the two sets of analytic charts are chosen to
agree (this would be the case if, for example, we divided the edge $e$ of section 3 into 2 pieces $e^1, e^2$
with $e=e^2\circ e^1$). In such a case the rightmost cell of the first edge coincides with the left most cell of
the second edge and hence the removal just removes this particular overcounting.}

\subsection{Sum to Product reformulation}\label{VII-2}
Equations (\ref{ap6-1}) and (\ref{2}) together with the fact that the number of edges $M$ is finite implies that 
\begin{equation}
V_{\triangle_{I_{p}}^{(p)}}\ =\ \int_{\triangle_{I_{p}}^{(p)}}N^{a}F_{a x_{p}}^{i}\tilde{E}^{x_{p}}_{i}\ +\ {O}(\delta v)
\end{equation}
where $O(\delta v)$ is bounded inpendent of $I_p, p$.
The vector constraint can be approximated as,
\begin{equation}
V[\vec{N}]\ =\ \sum_{p}\sum_{I_{p}}V_{\triangle_{I_{p}}^{(p)}}^{e_{p}}\ +\ \sum_{p}\sum_{I_{p}}\int_{\triangle_{I_{p}}^{(p)}}N^{a}F_{a\hat{b}}^{i}\tilde{E}^{\hat{b}}_{i}\ +\ \sum_{\triangle\in T^{* '}_{\gamma}}\int_{\triangle}N^{a}F_{ab}^{i}\tilde{E}^{b}_{i}\ +\ \textrm{O}(\delta)
\end{equation}

A proof almost identical to the one given in Appendix \ref{A1} yields $V[\vec{N}]$ in product form
\begin{equation}\label{27}
\begin{array}{lll}
V[\vec{N}]\ =\ \left\{\left(\prod_{p}\prod_{I_{p}}\left(1 + \frac{\delta}{(-il_{p}^{2}\gamma)} V_{\triangle_{I_{p}}^{(p)}}^{e_{p}}\right)\prod_{p}\prod_{I_{p}}\left(1 + \frac{\delta}{(-il_{p}^{2}\gamma)}\int_{\triangle_{I_{p}}^{(p)}}N^{a}F_{a\hat{b}}^{i}\tilde{E}^{\hat{b}}_{i}\right)\right.\right.\\
\vspace*{0.1in}
\hspace*{0.5in}\left.\left.\prod_{\triangle\in T^{'}_{\gamma}}\left(1 + \frac{\delta}{(-il_{p}^{2}\gamma)}\int_{\triangle}N^{a}F_{ab}^{i}\tilde{E}^{b}_{i}\right)\right)
-1\right\}\frac{(-il_{p}^{2}\gamma)}{\delta}\ +\ {O}(\delta)
\end{array}
\end{equation}

Thus a finite triangulation approximant, $V_{T}[\vec{N}]$, to $V[\vec{N}]$ can be defined, similar to (\ref{1}), as 
\begin{equation}\label{28}
\begin{array}{lll}
V_{T}[\vec{N}]\ =\ \left\{\left(\prod_{p}\prod_{I_{p}}\left(1 + \frac{\delta}{(-il_{p}^{2}\gamma)} V_{\triangle_{I_{p}}^{(p)}}^{e_{p}}\right)\prod_{p}\prod_{I_{p}}\left(1 + \frac{\delta}{(-il_{p}^{2}\gamma)}\int_{\triangle_{I_{p}}^{p}}N^{a}F_{a\hat{b}}^{i}\tilde{E}^{\hat{b}}_{i}\right)\right.\right.\\
\vspace*{0.1in}
\hspace*{0.5in}\left.\left.\prod_{\triangle\in T^{'}_{\gamma}}\left(1 + \frac{\delta}{(-il_{p}^{2}\gamma)}\int_{\triangle}N^{a}F_{ab}^{i}\tilde{E}^{b}_{i}\right)\right)
-1\right\}\frac{(-il_{p}^{2}\gamma)}{\delta}\ 
\end{array}
\end{equation}

Its operator correspondent is:
\begin{equation}\label{28a}
\begin{array}{lll}
\hat{V}_{T}[\vec{N}]\ =\ \left\{\left(\prod_{p}\prod_{I_{p}}\left(1 + \frac{\delta}{(-il_{p}^{2}\gamma)} \hat{V}_{\triangle_{I_{p}}^{(p)}}^{e_{p}}\right)\prod_{p}\prod_{I_{p}}\left(1 + \frac{\delta}{(-il_{p}^{2}\gamma)}\widehat{\int_{\triangle_{I_{p}}^{p}}N^{a}F_{a\hat{b}}^{i}\tilde{E}^{\hat{b}}_{i}}\right)\right.\right.\\
\vspace*{0.1in}
\hspace*{0.5in}\left.\left.\prod_{\triangle\in T^{'}_{\gamma}}\left(1 + \frac{\delta}{(-il_{p}^{2}\gamma)}\widehat{\int_{\triangle}N^{a}F_{ab}^{i}\tilde{E}^{b}_{i}}\right)\right)
-1\right\}\frac{(-il_{p}^{2}\gamma)}{\delta}\ 
\end{array}
\end{equation}

Consequently, a finite triangulation approximant ${ D}_{T}[\vec{N}]$ to ${ D}[\vec{N}]$ can be defined, 
similar to (\ref{2}), as.
\begin{equation}
\begin{array}{lll}
{ D}_{T}[\vec{N}]\ =\ \frac{-i\hbar}{\delta}\left\{\left[1 - \frac{\delta}{(-il_{p}^{2}\gamma)}{\cal G}_{T}[N^{i}]\right]\left[1 + \frac{\delta}{(-il_{p}^{2}\gamma}V_{T}[\vec{N}]\right]\right\} - 1
\end{array}
\end{equation}

An analysis along the lines of appendix \ref{A1} yields,
\begin{equation}
\left(1 - \frac{\delta}{(-il_{p}^{2}\gamma)}\hat{{\cal G}}_{T}[N^{i}]\right)\vert {\bf s}\rangle\ =\ \hat{{\cal U}}^{SU(2)}_{(N^{i},\delta)}\vert {\bf s}\rangle
\end{equation}
where, $\hat{{\cal U}}^{(SU(2)}_{(N^{i},\delta})$ is a  finite SU(2) gauge  transformation which acts 
in the standard way by on $\vert {\bf s}\rangle$ by rotating the intertwiners appropriately.

For each edge $e_{p}$ , we define $\hat{V}_{\triangle_{I_{p}}^{(p)}}$ as in 
section \ref{V}. 
Finally we order the product of operators in (\ref{28a}) 
exactly as written so that the contributions from the cells $\triangle\in T^{* '}_{\gamma}$ 
and the $\widehat{F_{a\hat{b}}^{i}\tilde{E}^{\hat{b}}_{i}}$ contributions of the cells $\triangle_{I_{p}}^{(p)}$
 are to the right. 
 Finally for each term (corresponding to each cell) in the product , we order the flux operator to the right.
With this choice of operator ordering it is easy to see that, just like in the case of a single edge, if 
$\triangle \in T^{*\prime}_{\gamma}$, 
\begin{equation}
\left(1 + \frac{\delta}{(-il_{p}^{2}\gamma)}\widehat{\int_{\triangle}N^{a}F_{ab}^{i}\tilde{E}^{b}_{i}}\right)\vert{\bf s}\rangle\ =\ \vert{\bf s}\rangle
\end{equation}
i.e.  the cells $\triangle\in T^{* '}_{\gamma}$ do not contribute anything as $T^{* '}_{\gamma}\cap\ \gamma\ =\ 0$.\\
Similar to the single edge case, we also have that
\begin{equation}
\left(1 + \frac{\delta}{(-il_{p}^{2}\gamma)}\widehat{\int_{\triangle_{I_{p}}^{p}}N^{a}F_{a\hat{b}}^{i}\tilde{E}^{\hat{b}}_{i}}\right)\vert{\bf s}\rangle\ =\ \vert{\bf s}\rangle
\end{equation}
because the edge $e_{p}$ is along the $x_{p}$ direction and $\hat{b}\ =\ \{y_{p},z_{p}\}$ whence 
$\hat{E}^{\hat{b}}_{i}\vert{\bf s}\rangle\ =\ 0.$\\
Hence just as for the single edge case((\ref{eq:3}) we have that 
\begin{equation}\label{29}
\left(1 + \frac{\delta}{-i\hbar}\hat{ D}_{T}[\vec{N}]\right)\vert{\bf s}\rangle\ =\ \hat{{\cal U}}^{SU(2)}_{(N^{i},\delta)}\left(\prod_{p}\prod_{I_{p}}\left(1 + \frac{\delta}{(-il_{p}^{2}\gamma)}\hat{V}^{e_{p}}_{\triangle_{I_{p}}^{(p)}}\right)\right)\vert{\bf s}\rangle
\end{equation}
Equation (\ref{29}) still has operator ordering issues we need to sort out.

Denote the edges of $\gamma$ which are along $\vec{N}$ by $e_{p_{\parallel}}^{(\parallel)}$ 
and those which are transverse to the integral curves of $\vec{N}$ ( except perhaps at their end points) by 
$e_{p_{\perp}}^{(\perp)}$ with $p_{\parallel}\ \in\ \{1,...,M_{\parallel}\}$ and $p_{\perp}\ \in\ \{1,...,M_{\perp}\}$, 
$M_{\parallel} + M_{\perp}\ =\ M$.
Let 
\begin{equation}\nonumber\\
\begin{array}{lll}
\cup_{p_{\perp}}e_{p_{\perp}}^{(\perp)}\ =:\ \gamma_{\perp}\\
\vspace*{0.1in}
\cup_{p_{\parallel}}e_{p_{\parallel}}^{(\parallel)}\ =:\ \gamma_{\parallel}
\end{array}
\end{equation}
Note that $\gamma_{\perp}$, $\gamma_{\parallel}$ are not necessarily connected graphs.

We order the right hand side of (\ref{29}) so that the contributions from edges in 
$\gamma_{\parallel}$ are to the right. 
Clearly, from the remarks below (\ref{2}) in section (\ref{V}), it follows that these contributions all reduce to unity. 
Whence we are only left with contributions coming from the edges of $\gamma_{\perp}$.\\

In order that the contributions from the
cells $\triangle_{I_{p_{\perp}}}^{p_{\perp}}$ and $\triangle_{I_{q_{\perp}}}^{q_{\perp}}$ 
act independently of each other  exactly in the manner described for a single edge, we need to ensure the following.
First, the surface   $S^{x_{p_{\perp}}}_{I_{p_{\perp}}}$ should intersect $\gamma$ only in the edge
$e_{I_{p_{\perp}}}^{p_{\perp}}$. This readily ensured by choosing $\delta_1, \delta_2$ to be sufficiently small.
Second, the surface $S^{x_{p_{\perp}}}_{I_{p_{\perp}}}$  should not intersect
any other edge which maybe generated by action of any $\hat{V}_{\triangle_{I_{q_{\perp}}}^{(q_{\perp})}}^{e_{q_{\perp}}}$.
To ensure this we need to slightly modify the choice of the surfaces labelling the electric flux variables in
equation (\ref{eq:ap4-1}) as follows.

Let 
\begin{equation}
\begin{array}{lll}
\phi(\vec{N},\delta)\circ e_{p_{\perp}}^{(\perp)}\ :=\ e_{p_{\perp} \delta}^{(\perp)}\\
\vspace*{0.1in}
\gamma_{\perp}(\delta)\ :=\ \cup_{p_{\perp}}\ e_{p_{\perp} \delta}^{(\perp)}
\end{array}
\end{equation}
From the application of  Appendix \ref{A4} to each pair of edges \\
\noindent
$\{\ (e_{p_{\perp} \delta}^{({\perp})}, e_{q_{\perp}}^{({\perp})})\ \vert\ \forall\ p_{\perp},q_{\perp}\}$ ,
it follows that 
we can find a small enough $\delta_{0}$ such that $\gamma_{\perp}\cap \gamma_{\perp \delta}$ is a 
finite set of isolated points in $\Sigma$ $\forall\ \delta\ \in\ (0,\delta_{0})$
\footnote{The cardinality of this set could depend
on $\delta$ but the important point is that for each such value of $\delta$, it is finite.}
Now for any $\delta\ \in\ (0,\delta_{0})$ we can proceed as follows. 

First, recall that the surface
$S^{x_{p_{\perp}}}_{I_{p_{\perp}}}$ is the rightmost face of the cell $\triangle_{I_{p_{\perp}}}$. However, 
equation (\ref{eq:ap4-1}) holds even if integrate over the electric flux through a surface 
$S^{(\epsilon )x_{p_{\perp}}}_{I_{p_{\perp}}}$ obtained by displacing $S^{x_{p_{\perp}}}_{I_{p_{\perp}}}$ to the left
by an amount $\epsilon << \delta$. We use this freedom to choose the surfaces so that they do not intersect
any of the finite number of isolated points in $\gamma_{\perp}\cap \gamma_{\perp}(\delta )$.
Hence if any face, $S^{x_{p_{\perp}}}_{I_{p_{\perp}}}$, intersects $\gamma_{\perp}\cap \gamma_{\perp}(\delta )$
in a point, we integrate the flux over the surface $S^{(\epsilon )x_{p_{\perp}}}_{I_{p_{\perp}}}$ instead.
In other words, 
just as we associated the loop $\gamma_I$ to the cell $\triangle_I$ 
in section 4, we associate the surface $S^{(\epsilon )x_{p_{\perp}}}_{I_{p_{\perp}}}$ to the cell 
$\triangle_{I_{p_{\perp}}}$ and use this surface (which is no longer the rightmost surface of the cell)
to evaluate the flux. 
The considerations of section 5 go through unchanged because the new flux operator still
intersects the edge $e_{p_{\perp}}^{({\perp})}$ away from the vertices of the triangulation 
$T_{e_{p_{\perp}}^{({\perp})}}$. In case the face $S^{x_{p_{\perp}}}_{I_{p_{\perp}}}$
does not intersect $\gamma_{\perp}\cap \gamma_{\perp}(\delta )$, we leave the choice of surface
unaltered. {\em From now on, we assume this choice has been made and in an abuse of notation,
continue to refer to the surfaces, whether displaced or not, by}  $\{S^{x_{p_{\perp}}}_{I_{p_{\perp}}}\}$.

Next, choose $\delta_{1}, \delta_{2}$ small enough that 
\begin{equation}
S_{I_{p_{\perp}}}^{x_{p_{\perp}}}\ \cap\ \gamma_{\perp}(\delta)\ =\ 0
\end{equation}
and such that 
$S_{I_{p_{\perp}}}^{x_{p_{\perp}}}$ intersects $\gamma$ only in the edge $e_{I_{p_{\perp}}}^{p_{\perp}}$.

With these choices,
each factor  in the product over edges in $\gamma_{\perp}$ acts independently exactly as in the case of a single edge.
\footnote{Note that we also need to ensure that the none of the surfaces, $S(I_{p_{\perp}}^{x_{p_{\perp}}})$ 
which label the fluxes intersect the
extra segments $s^{p_{\perp}}_{1}, s^{p_{\perp}}_{N_{p_{\perp}}}$ along the shift vector field. Since the shift
is transverse to the edges of $\gamma_{\perp}$ (except at perhaps a finite number of points), the edges of 
$\gamma_{\perp}$ can intersect these segments at most at a finite number of isolated points which can, once again,
be avoided by the surfaces $S(I_{p_{\perp}}^{x_{p_{\perp}}})$ by slightly moving them to the left as above.}
Whence in the present case each $h_{e_{p_{\perp}}}$ is mapped onto $h_{{{\bar e}_{p_{\perp}}}}$.\\
Finally the Gauss-law piece , $\hat{\cal U}^{(SU(2))}_{(\delta, N^{i})}$ removes the ``extra segments" from each 
${\bar e}_{p_{\perp}}$ as detailed in section (\ref{VI}). It also generates appropriate finite diffeomorphism on edges in $\gamma_{\parallel}$ , so that from (\ref{29}) we obtain,
\begin{equation}\label{ap28-1}
\left[1 + \frac{\delta}{-i\gamma l_{p}^{2}}\hat{{D}}_{T}[\vec{N}]\right]\vert{\bf s}\rangle\ =\ \hat{{\cal U}}(\phi(\delta, \vec{N}))\vert{\bf s}\rangle
\end{equation}
as required.\\
Note that if spin-network is gauge invariant, the Gauss law term acts as identity operator due to gauge invariance of vertex intertwiners. Thus as expected, we could have dropped 
${\cal G}[N^{i}]$ term from the classical expression and the accompanying heuristics of appendix \ref{A2} could have been avoided.

\section{The continuum limit on the LM habitat}
Let ${\cal D}$ be the finite span of spin network states and let 
${\cal D}^*$ be its algebraic dual. Let $[{\bf s}]$ be the set of spin networks
related by the action of diffeomorphism to $\bf s$. For each such 
diffeomorphism equivalence class of spin networks, $[{\bf s}]$ fix a 
``reference'' spin network ${\bf s}_0\in [{\bf s}]$. 
Let the vertices of (the coarsest graph underlying) ${\bf s}_0$ be denoted 
by ${\vec v}= (v_1,..,v_k)$ so that ${\vec v}\in \Sigma^{k({\bf s}_0)}$, 
$k({\bf s}_0)\equiv k$ being the number of vertices of ${\bf s}_0$.

Then the LM habitat,
${\cal D}^*_{LM}$, is defined as follows \cite{lm}. Let ${\cal D}^*_{LM}$ contain those
elements of  $\Psi$ of ${\cal D}^*$ for which 
\begin{equation}
\Psi (\vert {\bf s}\rangle )= \Psi_{{\bf s}_0}(\phi (v_1),.., \phi (v_k))
\label{defpsi}
\end{equation}
 where $\Psi_{{\bf s}_0}$ is a smooth complex valued function on 
$\Sigma^{k({\bf s}_0)}$ and $\phi$ is any diffeomorphism which maps
the reference spin network 
${\bf s}_0\in [{\bf s}] $ to ${\bf s}$ so that 
$(\phi (v_1),.., \phi (v_k))$ are the vertices of ${\bf s}$.
\footnote{Note that since $\Psi\in {\cal D}^*$, we have that 
$\Psi (\vert {\bf s}\rangle )= \Psi ({\hat U}_{\phi}\vert {\bf s}\rangle )$
for all diffeomorphisms $\phi$ which leave ${\bf s}$ invariant. Hence, if
there exist diffeomorphisms which preserve ${\bf s}$ but permute its vertices,
the function $\Psi_{{\bf s}_0}$ also has the property that it is invariant
under the corresponding permutation of its arguments.}

Thus, any $\Psi \in {\cal D}^*_{LM}$ is characterised by a family of smooth functions
$\Psi_{{\bf s}_0}$ one for each diffeomorphism class $[{\bf s}_0]$. 
These functions are referred to as {\em vertex smooth functions}\cite{lm}
Clearly, the set ${\cal D}_{diff}^*$ of
diffeomorphism {\em invariant} elements of ${\cal D}^*$ are obtained as those states $\Psi \in {\cal D}^*_{LM}$ 
for which each $\Psi_{{\bf s}_0}$ is a constant function. Thus ${\cal D}_{diff}^*\subset {\cal D}^*$.

The action of the diffeomorphism constraint operator  at finite triangulation (see equation (\ref{ap28-1})) on
$\Psi \in {\cal D}^*_{LM}$ is defined via its dual action:
\begin{eqnarray}
({\hat D}_T ({\vec N})\Psi )(\vert {\bf s}\rangle )&
=& \Psi( {\hat D}^{\dagger}_T ({\vec N})\vert {\bf s}\rangle)\\
&=&\frac{-i\hbar}{\delta} (\Psi(({\hat U}^{\dagger}_{\phi({\vec N},\delta )} - 1)\vert {\bf s}\rangle )\\
&=&\frac{-i\hbar}{\delta}(\Psi_{{\bf s}_0}( \phi^{-1}({\vec N},\delta )x_1,.,\phi^{-1}({\vec N},\delta )x_k)
                             -\Psi_{{\bf s}_0}(x_1,.,x_k))\nonumber\\
\end{eqnarray}
Here ${\bf s}_0$ is mapped to ${\bf s}$ by some diffeomorphism $\phi$ and $x_i= \phi  v_i$, where $v_i, i=1,..k$ are the vertices
of ${\bf s}_0$.

The action of the diffeomorphsim constraint operator, ${\hat D} ({\vec N})$,is obtained through the $\delta \rightarrow 0$
continuum limit of the action of its finite triangulation approximant ${\hat D}_T ({\vec N})$ so that:
\begin{eqnarray}
({\hat D}({\vec N})\Psi )(\vert {\bf s}\rangle )&=&
{-i\hbar}\lim_{\delta \rightarrow 0}
\frac{\Psi_{{\bf s}_0}( \phi^{-1}({\vec N},\delta )x_1,.,\phi^{-1}({\vec N},\delta )x_k)
                             -\Psi_{{\bf s}_0}(x_1,.,x_k))}{\delta} \nonumber\\
&=&i\hbar \sum_{i=1}^k N^a (x_i) \frac{\partial \Psi_{{\bf s}_0}(x_1,.,x_k))}{\partial x_i^a } 
\label{lm1}
\end{eqnarray}
where $x_i^a$ are the coordinates (in some coordinate chart) of the point $x_i\in \Sigma$.
Equation (\ref{lm1}) shows that the operator ${\hat D}({\vec N})$ is well defined on the habitat and maps the habitat 
state $\Psi$ specified by the family of  vertex smooth functions $\Psi_{{\bf s}_0}$ to the 
the habitat 
state $\Phi = {\hat D}({\vec N})\Psi$ specified by the family of  vertex smooth functions $\Phi_{{\bf s}_0}$ with 
\begin{equation}
\Phi_{{\bf s}_0}(v_{1},...,v_{k})= 
i\hbar \sum_{i=1}^k N^a (v_i) \frac{\partial \Psi_{{\bf s}_0}(v_1,.,v_k))}{\partial v_i^a } 
\label{lm2}
\end{equation}
From equation (\ref{lm2}) it follows that the joint kernel of the set of diffeomorpism constraint operators
$\{{\hat D}({\vec N}), \forall {\vec N}\}$ is the set of habitat states for which each vertex smooth function 
is a constant function. As mentioned above this set of states is precisely ${\cal D}_{diff}^*$.

From equations (\ref{lm1}) and (\ref{lm2}) it follows that 
\begin{equation}
{\hat D}({\vec N}){\hat D}({\vec M})- {\hat D}({\vec M}){\hat D}({\vec N})\Psi := \Phi^{{\vec N}, {\vec M}}
\end{equation}
with the habitat state $\Phi^{{\vec N}, {\vec M}}$ specified by the family of vertex smooth functions 
$\Phi^{{\vec N}, {\vec M}}_{{\bf s}_0}$ where
\begin{eqnarray}
\Phi^{{\vec N}, {\vec M}}_{{\bf s}_0}(x_{1},...,x_{k})&=&
(i\hbar )^2 \sum_{i=1}^k (M^a (x_i)\frac{\partial N^b(x_i)}{\partial x_i^a} 
-N^a (x_i)\frac{\partial M^b(x_i)}{\partial x_i^a}) 
\frac{\partial \Psi_{{\bf s}_0}(x_1,.,x_k))}{\partial x_i^b }\nonumber\\
&=& -(i\hbar )^2 \sum_{i=1}^k ({\cal L}_{\vec N}{\vec M})^b
\frac{\partial \Psi_{{\bf s}_0}(x_1,.,x_k)}{\partial x_i^b }.
\label{lm3}
\end{eqnarray}
Equation (\ref{lm3}) implies that, on ${\cal D}^*_{LM}$ we have that \
\begin{equation}
[{\hat D}({\vec N}),{\hat D}({\vec M})]= -i\hbar {\hat D}({\cal L}_{\vec N}{\vec M}),
\label{lm4}
\end{equation}
so that our construction of the diffeomorphism constraint operator results in an anomaly free (anti-)representation
of the Lie algebra of diffeomorphisms of $\Sigma$.

\section{Conclusions}
The diffeomorphism constraint $D({\vec N})$ generates diffeomorphisms along the
integral curves of the shift vector field ${\vec N}$. Hence, one expects
the quantum constraint operator, 
${\hat D}({\vec N})$, to have a non- trivial action at {\em all} the 
(infintely many) points  lying on those edges of a spin network state 
which are transverse  to ${\vec N}$.  In contrast, almost all operators of 
significance in LQG have a non- trivial action only at a {\em finite}
number of points namely the vertices of the graph underlying the spin network 
state. Indeed, the necessity of an action at infinitely many points was thought to be an obstacle 
to the construction of the operator ${\hat D}({\vec N})$ \cite{ttbook}. 
Our construction gets around this obstruction through the reformulation of the classical constraint
at finite triangulation as a {\em product} over 3- cells of the triangulation described in section 4.
This leads, in the quantum theory, to a {\em product} of bounded operators at finite triangulation
rather than a {\em sum}. The product admits a satisfactory continuum limit whereas the sum does not.
Thus, it is the passage to the product form which enables us to deal with the contributions from infinitely many
points in the continuum limit.

A sensible product reformulation also seems to require that the shift vector $N^a (x)$ cannot appear as an overall 
factor multiplying the diffeomorphism constraint $D_a (x)$ at the point $x$ because a product over all $x$ 
of shift vectors at each point $x$ is not an object which makes sense in the continuum limit. Hence it seems
inevitable that the shift vector dependence in $D({\vec N})$ at finite triangulation is taken care of by the
incorporation of both its direction and magnitude in the specification of the small loop which underlies the 
holonomy approximant to the Ashtekar- Barbero curvature, $F_{ab}^i$. Indeed, what we are able to construct is 
the quantity $N^aF_{ab}^i$ at finite triangulation rather than $F_{ab}^i$ itself. As a consequence, our construction
of curvature approximant bears a great conceptual similarity to that of Loop Quantum Cosmology (LQC) when viewed
in the following manner.

In isotropic LQC, the diffeomorphism constraint is satisfied identically and the Hamiltonian 
constraint reduces to its Euclidean part $H= \frac{\epsilon^{ijk}F_{abi}{\tilde E}^a_j{\tilde E}^b_k}{\sqrt{q}}$ where we
have used standard notation for the densitized triad and the determinant of the 3- metric. Our work here suggests that
rather than $F_{ab}^i$ it is $\frac{{\tilde E}^a_j}{\sqrt{q}} F_{ab}^i$ which needs to be approximated at finite
triangulation, and, that one should attempt to incorporate 
$\frac{{\tilde E}^a_j}{\sqrt{q}}$ as part of the specification of the small loop underlying the holonomy 
approximant. In the quantum theory such an attempt, if successful, would lead to the consideration of a loop
whose size depends on the triad operator, thus exhibiting a close conceptual similarity to the ``${\bar \mu}$''
scheme \cite{improvedlqc} for the Hamiltonian constraint in LQC.

Setting aside considerations of the Hamiltonian constraint, this work in itself (as seen in section 5)
reveals the necessity of a triad operator dependence in the construction of curvature approximants. This dependence
is both explicit (as seen in the occurence of the electric flux terms in equations (\ref{eq:(4)}) and (\ref{eq:16})) as well as
implicit in that the expressions for the curvature approximants depend on the spin label $j$ of the edge on which the 
curvature operator acts.
\footnote{Recall that the $j$ label specifies  the eigen values of the area operator which is built from the 
triads \cite{areaoprtr}.} A similar dependence of ``connection'' type operators on conjugate ``electric fluxes''
was also seen to be crucial in recent work on Polymer Parameterised Field Theory (PPFT) \cite{ppftham,ttham}.

Apart from this ``electric flux dependence'', one of the key lessons of our work in PPFT \cite{ppftham}
is the necessity of considering kinematically {\em singular} constraint operators 
in order to obtain a non- trivial representation of the constraint algebra.
Here, too, the existence of a non- trivial representation of the quantum constraint algebra can be traced to the 
kinematically singular nature of the diffeomorphism constraint operator. That this operator is singular on 
${\cal H}_{kin}$ is an obvious consequence of the factor of ${\delta}^{-1}$ in equation (\ref{final}).
It is this factor which leads to a non- trivial representation of the constraint algebra on the LM habitat in section 8.
Had this factor been absent the action of the constraint operator would have yielded the difference of the evaluations
of a vertex smooth function at points seperated by $\delta$.  This difference vanishes in the $\delta\rightarrow 0$
limit by virtue of the smoothness of the function. Instead, just as for PPFT \cite{ppftham}, the factor of 
$\delta^{-1}$ converts this difference into a derivative in the continuum limit, thus yielding a non- trivial 
action of the diffeomorphism constraint operator on the habitat as well as a non- trivial representation of the 
algebra of diffeomorphism constraints thereon.

Our final goal is the construction of the 
 Hamiltonian constraint operator in such a way as to obtain a non- trivial
anomaly free representation of its algebra. 
In the language of the concluding section of Reference 
\cite{ppftham}, let us refer to the quantum commutator between a pair of 
Hamiltonian constraints as
the Left Hand Side (LHS) and the  quantum correspondent of the classical Poisson bracket between this
pair as the Right Hand Side (RHS). The RHS is closely related to the diffeomorphism constraint operators
studied here; the only difference being that the shift vector field in the RHS is operator valued.
Earlier work by Thiemann \cite{qsd3}, Lewandowski and Marolf \cite{lm}
and Gambini, Lewandowski, Pullin and Marolf \cite{habitat2} showed that for density weight one Hamiltonian 
constraints, the algebra consistently trivialises i.e. the LHS and the RHS can be independently defined 
either with respect to the Uniform Rovelli- Smolin- Thiemann Topology on ${\cal H}_{kin}$ \cite{qsd3,ttbook} or
on the LM habitat \cite{lm,habitat2} and, in both cases,  both the RHS and the LHS vanish.
Our work on PPFT \cite{ppftham} (as well as the `rescaling by hand'
in Reference \cite{habitat2}) suggests the use of higher density weight constraints to probe the existence of
a {\em non- trivial} representation of the constraint algebra. Both these works also suggest that the 
current set of choices for curvature approximants are inappropriate. As emphasized in Reference \cite{habitat2}
the current set of choices used in the LHS do not result in an RHS which can move vertices by diffeomorphisms.
%In the language of the concluding section of Reference 
%\cite{mealokpft}, let us refer to the quantum commutator between a pair of (appropriately density weighted) 
%Hamiltonian constraints as
%the Left Hand Side (LHS) and the  quantum correspondent of the classical Poisson bracket between this
%pair as the Right Hand Side (RHS). 
%The evaluation of the LHS requires the construction of the Hamiltonian 
%constraint operator which, in turn, requires (among other things) 
%a suitable definition of the curvature, $F_{ab}^i$ at finite
%triangulation. The RHS is closely related to the diffeomorphism constraint operators studied here.
Since the choice of curvature approximants used in this work does result in the diffeomorphism constraint
moving vertices around by diffeomorphisms, 
the considerations of this work should be of use for a better understanding of both the LHS as well 
as the RHS. 

\noindent{\bf Acknowledgements}: AL would like to thank Miguel Campiglia, Adam Henderson and Casey 
Tomlin for discussions and Lois Sofia for help with the figures. MV thanks Fernando Barbero and Eduardo
Villase$\rm{\tilde n}$or for discussions. Work of AL is supported by NSF grant PHY-0854743 and by the Eberly Endowment fund.

\section*{Appendix}
\appendix

\section{Conventions}\label{A0}
In this appendix we summarise various conventions used in the computations. Our conventions are same as those given in \cite{ttbook}

\begin{equation}
\begin{array}{lll}
\tau^{i} = -i\sigma^{i}\\
\vspace*{0.1in}
h_{e}(A)\ =\ {\cal P}\exp\left[\int_{e}\frac{A_{i}\tau^{i}}{2}\right]\\
\end{array}
\end{equation}
where $\{\sigma^{i}\}$ are  the Pauli matrices.
 
e.g. above conventions imply that, 

\begin{equation}
h_{\alpha}\ =\ 1\ +\ \frac{\delta^{2}}{2}F^{i}\tau_{i}\ +\ \textrm{O}(\delta^{3})
\end{equation}
for any plaquette $\alpha$  of co-ordinate area $\delta^{2}$.

\section{Proof of Equation (\ref{eq:ap3-2})}\label{A1}
It is straightforward to see that equation (\ref{eq:ap3-2}) can be rewritten in the for,m:
\begin{equation}\label{eq:A.1.1}
\begin{array}{lll}
\left\{\prod_{I}\left(1 + \frac{\delta V_{\triangle_{I}}^{(e)}}{-il_{p}^{2}\gamma}\right)\prod_{J}\left(1 +\frac{\delta}{-il_{p}^{2}\gamma}\int_{\triangle_{J}}N^{a}F_{a\hat{b}}^{i}\tilde{E}^{b}_{i}\right)\prod_{\triangle\notin T^{*}_{e}}\left(1 + \frac{\delta}{-il_{p}^{2}}\int_{\triangle}N^{a}F_{ab}^{i}\tilde{E}^{b}_{i}\right)\right\}\\
\vspace*{0.1in}
-\left\{ 1\ +\ \frac{\delta}{-i l_{p}^{2}\gamma}\left(\sum_{I}V_{\triangle_{I}}^{(e)}\ +\ \sum_{I}\int_{\triangle_{I}}N^{a}F_{a\hat{b}}^{i}\tilde{E}^{\hat{b}}_{i}\ +\ \sum_{\triangle\in T_{e}^{*}}\int_{\triangle}N^{a}F_{ab}^{i}\tilde{E}^{b}_{i}
\right)\right\} 
=\ O(\delta^{2})
\end{array}
\end{equation}
We now prove equation (\ref{eq:A.1.1}).

%Also let the volume of $\Sigma$ as measured by the three form $\omega$ be V whence
%\begin{equation}\label{eq:0.2}
%\sum_{\triangle\in T^{*}}v\ =\ V
%\end{equation}

%{\bf Part 1} : To show that $\vert \textrm{R.H.S.} - \textrm{L.H.S.}\vert\ <\ \textrm{Analytic function in $\delta$}$.\\
\noindent{\bf Proof} : \\
Note that $N^a, A_a^i, {\tilde E}^b_i$ and the volume form $\omega$ (see section 3) 
are smooth tensor fields on $\Sigma$ and that any smooth function on $\Sigma$ is 
bounded  by virtue of the compactness of $\Sigma$. This, in conjunction with equations (\ref{vol=v}),
(\ref{triangleibound}) and (\ref{vtrianglei}) imply that there exists some positive constant ${\cal C}$ 
which is independent
of $\delta, \delta_1,\delta_2,I$ such that
\begin{equation}\label{defcalc}
\begin{array}{lll}
\vert \int_{\triangle_{I}}N^{a}F_{a\hat{b}}^{i}\tilde{E}^{\hat{b}} < {\cal C}v\\
\vspace*{0.1in}
\vert \int_{\triangle} N^{a}F_{ab}^{i}\tilde{E}^{b}_{i}\vert\ <\ {\cal C}v\\
\vspace*{0.1in}
\vert V_{\triangle_{I}}^{(e)}\vert\ <\ {\cal C}v
\end{array}
\end{equation}

Let us refer to the Left Hand Side of equation (\ref{eq:A.1.1})  by the abbreviation $L.H.S.$ 
Expanding the first set of terms (in curly brackets) 
of the $L.H.S.$ and using the bounds (\ref{defcalc}), we obtain
\begin{equation}
\begin{array}{lll}
\vert L.H.S.\vert\ <\\
\vspace*{0.1in}
\hspace*{0.6in} \left[\prod_{I}\left(1 + \frac{\delta}{l_{p}^{2}\gamma} {\cal C}v\right)\right]^{2}\prod_{\triangle\notin T_{e}^{*}}\left(1 + \frac{\delta}{l_{p}^{2}\gamma}{\cal C}v\right)
-\ 2\sum_{I}\frac{\delta}{l_{p}^{2}\gamma}{\cal C}v - \sum_{\triangle\notin T^{*}_{e}}\frac{\delta}{l_{p}^{2}\gamma}{\cal C}v -\ 1
\end{array}
\label{lhs}
\end{equation}

Next, denote the volume of $\Sigma$ (as measured by $\omega$) by V so that $V=\int_{\Sigma}\omega$. From 
equation (\ref{vol=v}) we have that 
\begin{equation}
V= \sum_{I=1}^N\int_{\triangle_I} \omega + \sum_{\triangle\notin T^{*}_e}v
\label{totalv}
\end{equation}
For sufficiently small $\delta$, equation (\ref{eq:0.1}) implies that 
\begin{equation}
\sum_{I=1}^N\int_{\triangle_I} \omega < \sum_{I=1}^N Dv
<\frac{2L}{\delta}Dv = O(\delta_1\delta_2) .
\label{volte}
\end{equation}
Equations (\ref{totalv}), (\ref{volte}) imply that 
\begin{equation}
\sum_{\triangle\notin T^{*}_{e}}v = V+ O (\delta_1\delta_2)
\label{sumnotintev}
\end{equation}
and equation (\ref{volte}) implies that 
\begin{equation}
\sum_{I=1}^N v = O(\delta_1\delta_2)
\label{sumiv}
\end{equation}
Using equations (\ref{sumnotintev}), (\ref{sumiv}) in the summations in equation (\ref{lhs}) 
%and using the fact that $(\cup_{\triangle\notin T^*_e}\triangle ) \subset T^*$ in the second product of 
%equation (\ref{lhs})
yields
\begin{equation}\label{eq:A.1.3}
\vert L.H.S.\vert <\  \prod_{I}\left(1 + \frac{\delta}{l_{p}^{2}\gamma}{\cal C}v\right)^2
\prod_{\triangle\notin T^{*}_e}\left(1 + \frac{\delta}{l_{p}^{2}\gamma}{\cal C}v\right)\ 
- V\frac{{\cal C}\delta}{l_{p}^{2}\gamma} -1 +O(\delta_1\delta_2).
\end{equation}

Next, consider the first product in the above equation. For $\delta_1, \delta_2$ sufficiently smaller than $\delta$,
we have that 
%\begin{equation}\label{eq:A.1.4.}
%\delta_{1}\delta_{2} = O(\delta^{2})
%\end{equation}
\begin{equation}\label{eq:ap2-1}
\begin{array}{lll}
\prod_{I}(1 + \frac{\delta{\cal C}v}{l_{p}^{2}\gamma})^2 <\ (1 + \frac{\delta{\cal C}v}{l_{p}^{2}\gamma})^{\frac{L}{\delta_{1}}\frac{L}{\delta_{2}}}\\
\vspace*{0.1in}
\hspace*{1.0in}=\ (1 + \frac{\delta^{2}{\cal C}\delta_{1}\delta_{2}}{l_{p}^{2}\gamma})^{\frac{L^{2}}{\delta_{1}\delta_{2}}}\\
\vspace*{0.1in}
\hspace*{1.0in}=\ (1 + \frac{\delta^{2}{\cal C}\delta_{1}\delta_{2}}{l_{p}^{2}\gamma})^{\frac{1}{\frac{\delta_{1}\delta_{2}{\cal C}\delta^{2}}{l_{p}^{2}\gamma}}L^{2}{\cal C}\frac{\delta^{2}}{l_{p}^{2}\gamma}}
\end{array}
\end{equation}
where the first line follows from equation (\ref{eq:0.1}) and $\delta_1,\delta_2<<\delta$ and the 
second line uses $v=\delta\delta_1\delta_2$. 

In order to estimate the third line of the above equation
we use the identity
$\lim_{x\rightarrow 0}(1 + x)^{\frac{a}{x}}\ =\ e^{a}$.
This identity implies that for any $\epsilon>0$ there exists small enough $x_0>0$ such that for all $x$ with  $0<x<x_0$,
we have that $\vert (1 + x)^{\frac{a}{x}}- e^a| <\epsilon$. 
We set $\epsilon = \delta^2$, $a=\frac{\delta^2{\cal C}L^2}{l_P^2\gamma}$ and 
$\frac{{\cal C}\delta^2\delta_1\delta_2}{l_P^2\gamma} =x$. Then for a given $\delta$, 
we can always choose $\delta_1,\delta_2$ small enough so as to obtain
\begin{equation}\label{eq:ap2-2}
\begin{array}{lll}
\left(1 + \delta^{2}\frac{{\cal C}\delta_{1}\delta_{2}}{l_{p}^{2}\gamma}\right)^{\frac{1}{\delta_{1}\delta_{2}\frac{{\cal C}\delta^{2}}{l_{p}^{2}\gamma}}\frac{L^{2}{\cal C}\delta^{2}}{l_{p}^{2}\gamma}}\ =\ e^{\frac{L^{2}{\cal C}\delta^{2}}{l_{p}^{2}\gamma}} + \textrm{O}(\delta^{2})\\
\vspace*{0.1in}
\hspace*{0.3in} = 1 + \textrm{O}(\delta^{2})
\end{array}
\end{equation}
Equations (\ref{eq:ap2-1}) and (\ref{eq:ap2-2}) imply that
\begin{equation}\label{eq:A.1.5}
\prod_{I}\left(1 + \frac{\delta{\cal C}v}{l_{p}^{2}\gamma}\right)\ <\ 1 + \textrm{O}(\delta^{2})
\end{equation}

Finally, consider the second product 
%$\prod_{\triangle\notin T^{*}_e}\left(1 + \frac{\delta^{2}}{l_{p}^{2}\gamma}{\cal C}v\right)$ 
in (\ref{eq:A.1.3}).
We have that 
\begin{equation}
\begin{array}{lll}
\prod_{\triangle\notin T^{*}_e}\left( 1 + \frac{\delta{\cal C}v}{l_{p}^{2}\gamma}\right)\ <\ \left(1 + \frac{\delta{\cal C}v}{l_{p}^{2}\gamma}\right)^{\frac{V}{v}}\\
\vspace*{0.1in}
\hspace*{0.8in}=\ \left(1 + \frac{\delta{\cal C}v}{l_{p}^{2}\gamma}\right)^{\frac{V}{\frac{\delta{\cal C}v}{l_{p}^{2}\gamma}}\frac{\delta{\cal C}}{l_{p}^{2}\gamma}}
\end{array}
\end{equation}
where we have used equation (\ref{totalv}) in the first line. Once again, 
given $\delta$, for small enough $\delta_{1},\delta_{2}$ we have that 
\begin{equation}\label{eq:A.1.6}
\left(1 + \frac{\delta{\cal C}v}{l_{p}^{2}\gamma}\right)^{\frac{1}{\frac{\delta{\cal C}v}{l_{p}^{2}\gamma}}V\frac{\delta{\cal C}}{l_{p}^{2}\gamma}}\ =\ e^{\frac{\delta V{\cal C}}{l_{p}^{2}\gamma}} + \textrm{O}(\delta^{2})
\end{equation}

Using equations (\ref{eq:A.1.5}), (\ref{eq:A.1.6}) in equation (\ref{eq:A.1.3}) together with the fact that 
$\delta_1,\delta_2<<\delta$, we have that
\begin{equation}
\begin{array}{lll}
\vert L.H.S.\vert\ <\ \left[1 + \textrm{O}(\delta^{2})\right]\left[e^{\frac{\delta V{\cal C}}{l_{p}^{2}\gamma}}\ +\ \textrm{O}(\delta^{2})\right] + \textrm{O}(\delta^{2})\ -\ \frac{V{\cal C}\delta}{l_{p}^{2}\gamma} -1\\
\vspace*{0.1in}
\hspace*{0.4in}=\ \left[1 + \textrm{O}(\delta^{2})\right]\left[1 + \frac{\delta V{\cal C}}{l_{p}^{2}\gamma} + \textrm{O}(\delta^{2})\right]\ -\ \frac{V{\cal C}\delta}{l_{p}^{2}\gamma} - 1 + \textrm{O}(\delta^{2})\\
\vspace*{0.1in}
\hspace*{0.4in}=\ \textrm{O}(\delta^{2})
\end{array}
\end{equation}
This completes the proof.

%\pagebreak
\section{Quantization of the Gauss term}\label{A2}

In this appendix we quantize the gauss part of the Diffeomorphism constraint.\\

Classically we have
\begin{equation}
{\cal G}[N^{i}]\ =\ \int_{\Sigma} N^{i}{\cal D}_{a}\tilde{E}^{a}_{i}
\end{equation}
where $N^{i}\ =\ N^{a}\cdot A_{a}^{i}$.\\
We shall treat this quantization a bit heuristically. This is because, \\
\noindent {(i)} Our main focus is on the approximants to $F_{ab}^{i}$ and \\
\noindent {(ii)} As we shall see in section (\ref{VII}), for gauge invariant states we can drop the $\hat{{\cal G}}[N^{i}]$ term altogether and start off from a classical expression wherein the Gauss constraint is already imposed.\\
Our treatment will be similar to that for deriving the action of flux operators in \cite{areaoprtr}. Whence we will first set $\hat{E}^{a}_{i}\ =\ \frac{\hbar}{i}G\gamma \frac{\delta}{\delta A_{a}^{i}}$ and restrict attention to cylindrical functions of \emph{smooth connections}. This will result in an operator whose action can be naturally generalised to functions of generalised connections.\\
Note that
\begin{equation}
\begin{array}{lll}
-{\cal G}[N^{i}]\ =\ -\int_{\Sigma}N^{i}{\cal D}_{a}\tilde{E}^{a}_{i}\ =\ \int_{\Sigma}\left({\cal D}_{a}N^{i}\right)\tilde{E}^{a}_{i}
\end{array}
\end{equation}
Let $h_{e}(A)$ be the holonomy of a smooth connection $A$. In what follows, terms of $\textrm{O}(\delta^{n})$ will be defined in the context of this fixed connection $A$.  i.e. 
\begin{equation}
\begin{array}{lll}
\alpha(A,\delta)\ =\ \textrm{O}(\delta^{n})\ <=>\  \lim_{\delta\rightarrow 0,\ A\ \textrm{fixed}}\ \frac{\alpha(A,\delta)}{\delta^{n}}\ \textrm{exists}
\end{array}
\end{equation}

Then, we have the following exact result \cite{ttbook},

\begin{equation}
\begin{array}{lll}
-\frac{\delta}{-il_{p}^{2}\gamma}\hat{\cal G}[N^{i}]h_{e}(A)\ =\ \left[-\left(\delta N(v_{1})\right)h_{e}(A)\ +\ h_{e}(A)\left(\delta N(v_{N})\right)\right]
\end{array}
\end{equation}
where $N(v)\ :=\ \frac{N^{i}\tau_{i}}{2}$ and where $v_{1}, v_{N}$ are the beginning and end points of $e$ as in the main text.

%\begin{equation}\label{ap24-1}
%\begin{array}{lll}
%-\frac{\delta}{-il_{p}^{2}\gamma}\hat{\cal G}[N^{i}]h_{e}(A)\ =\ \frac{\delta}{(-il_{p}^{2}\gamma)}\int_{\Sigma}\left({\cal D}_{a}N^{i}\right)\left(-i\hbar G\gamma\right)\frac{\delta}{\delta A_{a}^{i}}h_{e}(A)\\
%\vspace*{0.1in}
%\hspace*{0.5in} =\ \int_{\Sigma}\left(\delta {\cal D}_{a}N^{i}\right)\frac{\delta}{\delta A_{a}^{i}}h_{e}(A)\\
%\vspace*{0.1in}
%\hspace*{0.5in} =\  h_{e}(A_{a}^{i} + \delta{\cal D}_{a}N^{i})\ -\ h_{e}(A_{a}^{i})\ +\ \textrm{O}(\delta^{2})
%\end{array}
%\end{equation}

%As $N^{i}\ =\ N^{a}\cdot A_{a}^{i}$ is  $su(2)$ valued scalar-field, we can define; via the exponential mapping

%\begin{equation}
%g(\delta N^{i}(x))\ :=\ \exp[\delta N^{i}(x)]
%\end{equation}

%Using which we can re-write the R.H.S of (\ref{ap24-1}) as,

%\begin{equation}\label{ap25-1}
%\begin{array}{lll}
%-\frac{\delta}{-il_{p}^{2}\gamma}\hat{\cal G}[N^{i}]h_{e}(A)\ =\ g(\delta N^{i}(v_{1}))^{-1}\ h_{e}(A)\ g(\delta N^{i}(v_{N})) - h_{e}(A)\ +\ \textrm{O}(\delta^{2})
%\end{array}
%\end{equation}

%As is well known (\cite{ttbook}),
%the Hamiltonian vector field $-\nu_{{\cal D} N^{i}}$ of the Gauss constraint  ${\cal G}[N^{i}]\ =\ -E_{i}[{\cal D}N^{i}]$  on (classical) holonomy functionals is given by,
%\begin{equation}\label{eq:ap5-1}
%\left(-\nu_{{\cal D} N^{i}}h_{e}\right)(A)\ =\ \left(\gamma G\right) \left[\left(h_{e}(A)\circ(N^{i}(f(e))\tau_{i})\right) - \left(N^{i}(b(e))\tau_{i})\circ h_{e}(A)\right)\right]
%\end{equation}
%where $b(e)$ and $f(e)$ are the beginning and final points of the edge $e$ respectively.\\
Next consider the segments $s_{1}$ and $s_{N}$ which are of affine length $\delta$ along the integral curves of $\vec{N}$ emanating from $v_{1}$ , $v_{N}$ respectively.\\
Clearly,
\begin{equation}
\begin{array}{lll}
\frac{h_{s_{1}^{-1}}(A) - 1}{\delta}\ =\ -N^{i}(v_{1})\frac{\tau_i}{2}\ +\ \textrm{O}(\delta)\\
\vspace*{0.1in}
\frac{h_{s_{N}}(A)\ - 1}{\delta}\ =\ N^{i}(v_{N})\frac{\tau_i}{2}\ +\ \textrm{O}(\delta)
\end{array}
\end{equation}

Whence,

%\begin{equation}\label{ap25-2}
%\begin{array}{lll}
%h_{s_{1}}(A)\ =\ g(N^{i}(v_{1}))\ +\ \textrm{O}(\delta^2)\\
%\vspace*{0.1in}
%h_{s_{N}}(A)\ =\ g(N^{i}(v_{N}))\ +\ \textrm{O}(\delta^2)
%\end{array}
%\end{equation}

\begin{equation}\label{ap25-2}
\begin{array}{lll}
-\frac{\delta}{-il_{p}^{2}\gamma}\hat{\cal G}[N^{i}]h_{e}(A)\ =\ \left[h_{s_{1}^{-1}}\circ h_{e}\circ h_{s_{N}}\ -\ h_{e}\right]\ +\ \textrm{O}(\delta^{2})
\end{array}
\end{equation}

Thus we define the action of an ``approximant" $\hat{{\cal G}}_{T}[N^{i}]$ to $\hat{{\cal G}}[N^{i}]$ (derived above) on cylindrical functions of smooth connections as
\begin{equation}
\frac{-\delta}{-il_{p}^{2}\gamma}\hat{{\cal G}}_{T}[N^{i}]\ h_{e}\ =\ h_{s_{1}}^{-1}\cdot h_{e}\cdot h_{s_{N}}\ -\ h_{e}
\end{equation}
and assume the same action on cylindrical functions of the generalised connections.

Whence,
\begin{equation}\label{ap9-1}
\left(1 - \frac{\delta}{-il_{p}^{2}\gamma}\hat{{\cal G}}_{T}[N^{i}]\right) (h_{e})_{A}^{\;\;\;\;B}\ =\ (h_{s_{1}}^{-1}\cdot h_{e}\cdot h_{s_{N}})_{A}^{\;\;\;\;B}
\end{equation}

{\bf Corollary} : \\
If $\vec{N}$ is tangential to $e$, $\hat{{\cal G}}_{T}[N^{i}]$ acts via spatial diffeomorphism of the edge $e$.\\
{\bf Proof} : \\
if $\vec{N}$ is tangential to $e$, then (as shown in figure A.1)
\begin{equation}
h_{s_{1}}^{-1}\cdot h_{e}\cdot h_{s_{N}}\ =\ h_{\Phi(\vec{N},\delta)\circ e}
\end{equation}

q.e.d
\psfrag{a}{$s_{1}$}
\psfrag{b}{$\vec{N}$}
\psfrag{c}{$s_{N}$}
\psfrag{d}{$s_{1}$}
\psfrag{e}{$s_{N}$}
\begin{center}
\includegraphics[scale=0.7]{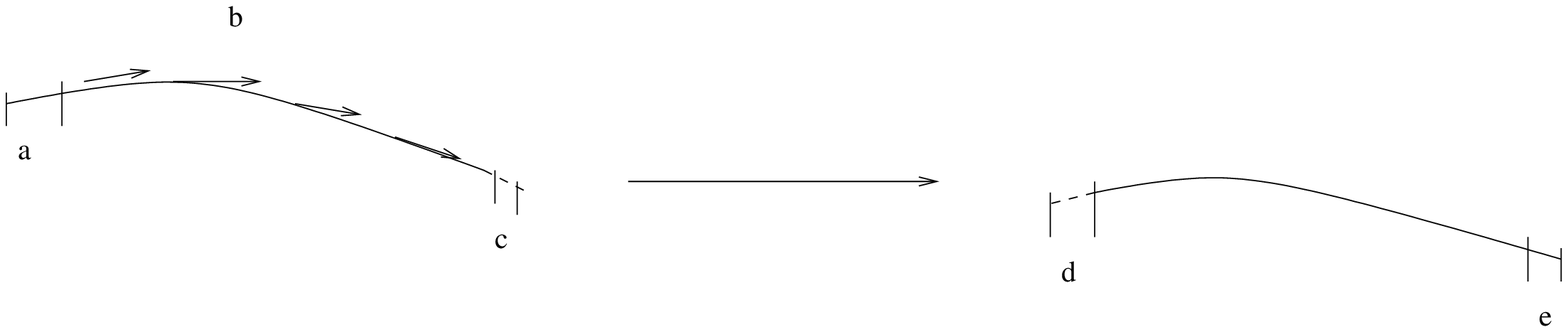}
Fig A.1
\end{center}

\section{Holonomy around infinitesimal loop}\label{A3}

In this section we show that leading order term in the expansion of $\textrm{Tr}(h_{\gamma_{I}}\tau^{i})$ is precisely $N^{a}F_{ax}^{i}\delta^{2}$.\\
For the benefit of the readers, we recall certain structures introduced in the main text which would be needed in this section.\\
\noindent {\bf (1)} $\gamma_{I}$ is a loop formed by $e_{I}$ and $\overline{e}_{I} := \phi(\vec{N},\delta)\circ e_{I}$ with the remaining two sides of the loop obtained by joining $b(e_{I})$ with $b(\overline{e}_{I})$ and $f(e_{I})$ with $f(\overline{e}_{I})$ respectively.\\
\noindent {\bf (ii)} The y-z co-ordinates are chosen so that $N^{y}, N^{z}$ are positive semi-definite, which implies that $\gamma_{I}$ is transversed clockwise.\\

Now the standard results for the expansion of small loop holonomy yields \footnote{our conventions are stated in appendix \ref{A0}.}
\begin{equation}
-Tr(h_{\gamma_{I}}\tau^{i})\ =\ -F_{ab}(v)\epsilon^{ab}(v)\int_{S(\gamma_{I})}\epsilon_{ab}\ +\ \textrm{O}(\delta^{3})
\end{equation}
where $S(\gamma_{I})$ is the surface spanned by $\gamma_{I}$, $\epsilon_{ab}$ is some co-ordinate two form on $S(\gamma_{I})$ and $\epsilon^{ab}$ is it's inverse.
On choosing $S(\gamma_{I})$ to be the open surface in the interior of $\gamma_{I}$ (with it's boundary being $\gamma_{I}$), we have,
\begin{equation}
\begin{array}{lll}
\epsilon_{ab}\ =\ dx\wedge d\lambda\\
\vspace*{0.1in}
\frac{1}{2}F_{ab}^{i}\epsilon^{ab}\ =\ F_{x\lambda}\ =\ -N^{a}F_{ax}^{i}\\
\vspace*{0.1in}
\int_{S(\gamma_{I})}\epsilon_{ab}\ =\ \delta^{2}
\end{array}
\end{equation}
Whence,
\begin{equation}
\textrm{Tr}(h_{\gamma_{I}}\tau^{i})\ =\ \delta^{2}N^{a}F_{ax}^{i}\ + \textrm{O}(\delta^{3})
\end{equation}

\section{Intersection of analytic edges}\label{A4}

\noindent 
{\bf Lemma 1}: Let $e_1,e_2$ be compact, connected, non- self intersecting 
 analytic edges which are 
analytically extendable past their endpoints. Let $e_1, e_2$ intersect
in a single point $v$. Let $\phi({\vec N},\delta)$ be as in the main text,
with ${\vec N}$ transverse to $e_1,e_2$ except perhaps at their end points.
Let $\phi({\vec N},\delta)\circ e_1:=e_1 (\delta ).$
Then there exists $\delta_0$ such that for each $\delta <\delta_0$, we have
that $e_1(\delta) \cap e_2$ consists of a finite number of 
isolated intersection points.

\noindent{\bf Proof}: Let the other end point of $e_1$ be $v^{\prime}$.
Lets assume the contrary i.e. for any $\delta_0>0$ there exist infinitely many
$\delta <\delta_0$  s.t. 
$e_1(\delta) \cap e_2$ is not a finite number of isolated points.
Thus, there exist infinitely many $\delta$ in any open neighbourhood of 
$\delta =0$ such that 
$e_1(\delta) \cap e_2$ contains a closed segment of  $e_1(\delta)$.

Since $e_2$ is connected and analytic, either \\
\noindent (i) $e_2\subset e_1 (\delta )$ 
for infinitely many $\delta$ close to 
zero, or\\
\noindent (i)$\phi({\vec N},\delta)\circ v^{\prime} \in e_2$  
for infinitely many $\delta$ close to zero, or\\
\noindent (iii) $\phi({\vec N},\delta)\circ v \in e_2$  
for infinitely many $\delta$ close to zero.

Case (i) is impossible due to the transversality of ${\vec N}$ with respect
to $e_1$. To see this, let $p$ be a point in the interior of $e_1$
Transversality implies that for any such point $p$, there exists 
$\epsilon_0(p)$ such that for every 
$\epsilon$ with $0<\epsilon <\epsilon_0(p)$, we have that 
\begin{equation}
\phi({\vec N}, \epsilon )\circ p := p_{\epsilon} \notin e_1 .
\label{transverse}
\end{equation}
On the other hand if case (i) is true then there exist 
$\delta_1,\delta_2$
with $0<\delta_1<\delta_2<\epsilon_0(p)$, $q\in e_2$, $p\in {\rm Int}e_1$, $p^{\prime}\in e_1$ such that 
$\phi({\vec N}, \delta_1)\circ p^{\prime}= 
\phi({\vec N}, \delta_2)\circ p= q$
so that $\phi({\vec N}, \delta_2- \delta_1)\circ p = p^{\prime}\in e_1$ 
which is in 
contradiction with condition (\ref{transverse}) above.

Case (ii) is impossible as the sequence $\phi({\vec N},\delta)\circ v^{\prime}$
converges to $v^{\prime}$ as $\delta$ decreases and $v^{\prime}\notin e_2$
which contradicts the compactness of $e_2$.

In case (iii) if ${\vec N}$ is non- vanishing at $v$, the 
 sequence $\phi({\vec N},\delta)\circ v$
converges to $v$ along the integral curve of ${\vec N}$
which contradicts the transversality of $N^a$ with respect to $e_2$.
Hence ${\vec N}$ must vanish at $v$. But then it must be the case that 
either $e_2\subset e_1(\delta )$  for infinitely many $\delta$ close to zero
or $\phi({\vec N},\delta)\circ v^{\prime} \in e_2$  
for infinitely many $\delta$ close to zero. These are cases (i) and (ii)
which we have shown to be impossible.

This completes the proof.\\

\noindent
{\bf Lemma 2}: 
Let $e_1,e_2$ be compact, connected, non- self intersecting 
 analytic edges which are 
analytically extendable past their endpoints. Let $e_1 \cap e_2$ be empty.
Let $\phi({\vec N},\delta)$ be as in the main text,
with ${\vec N}$ transverse to $e_1,e_2$ except perhaps at their end points.
Let $\phi({\vec N},\delta)\circ e_1:=e_1 (\delta ).$
Then there exists $\delta_0$ such that for each $\delta <\delta_0$, we have
that $e_1(\delta) \cap e_2$ consists of a finite number of 
isolated intersection points.

\noindent{\bf Proof}: 
We have either case (i) or case (ii) with $v^{\prime}$ being either of 
the end points of $e_1$ and the proof of impossibility of these cases is 
identical to that in Lemma 1.\\

\noindent{\bf Note}: We believe that the following stronger statement holds:\\
There is some $\delta_0$
such that for all $\delta$ with $0<\delta <\delta_0$, we have that 
$e_1 (\delta )\cap e_2$ is empty.

However, since Lemma 2 is sufficient for our purposes we will not 
attempt to prove the stronger statement here.

%\psfrag{e}{$e$}
%\psfrag{f}{$\phi(\vec{N},\delta)$}
%\psfrag{g}{$\phi(\vec{N},\delta)\circ e$}

%\begin{center}
%\includegraphics[scale=0.7]{fig-a.eps}
%fig-a
%\end{center}

%\psfrag{m}{$e$}
%\psfrag{n}{$\phi(\vec{N},\delta)$}
%\psfrag{p}{$\overline{e}(\vec{N},\delta)$}

%\begin{center}
%\includegraphics[scale=0.7]{fig-b.eps}
%fig-b
%\end{center}

%\psfrag{r}{$\overline{e}(\vec{N},\delta)$}
%\psfrag{s}{$\phi(\vec{N},\delta)\circ e$}

%\begin{center}
%\includegraphics[scale=0.7]{figurec.eps}
%fig-c
%\end{center}

%\psfrag{u}{$e$}
%\psfrag{vI}{$v_{I}$}
%\psfrag{vI+1}{$v_{I+1}$}
%\psfrag{ebar}{$\overline{e}$}
%\begin{center}
%\includegraphics[scale=0.7]{figured.eps}
%fig-d
%\end{center}

%\psfrag{x}{$\overline{e}(\vec{N},\delta)$}

%\begin{center}
%\includegraphics[scale=0.7]{figuree.eps}
%fig-e
%\end{center}

\end{document}